Check for updates

# Can learning from natural image denoising be used for seismic data interpolation?

Hao Zhang[1], Xiuyan Yang[2], and Jianwei Ma[2]

## ABSTRACT

We have developed an interpolation method based on the denoising convolutional neural network (CNN) for seismic data. It provides a simple and efficient way to break through the problem of the scarcity of geophysical training labels that are often required by deep learning methods. This new method consists of two steps: (1) training a set of CNN denoisers to learn denoising from natural image noisy-clean pairs and (2) integrating the trained CNN denoisers into the project onto convex set (POCS) framework to perform seismic data interpolation. We call it the CNN-POCS method. This method alleviates the demands of seismic data that require shared similar features in the applications of end-to-end deep learning for seismic data interpolation. Additionally, the adopted method is flexible and applicable for different types of missing traces because the missing or downsampling locations are not involved in the training step; thus, it is of a plug-and-play nature. These indicate the high generalizability of the proposed method and a reduction in the necessity of problem-specific training. The primary results of synthetic and field data show promising interpolation performances of the adopted CNN-POCS method in terms of the signal-to-noise ratio, dealiasing, and weak-feature reconstruction, in comparison with the traditional $f$-$x$ prediction filtering, curvelet transform, and block-matching 3D filtering methods.

## INTRODUCTION

Due to existing terrain obstacles or economic restrictions, missing traces in acquired seismic data, nonuniformly or uniformly,

along the spatial coordinate is unavoidable, and this affect seismic inversion, amplitude-versus-angle analysis, and migration. To use these incomplete data, many researchers have developed dozens of interpolation methods to restore the missing traces. Besides frequency-space ($f$-$x$) prediction filtering methods (Spitz, 1991; Naghizadeh and Sacchi, 2009), other methods based on the sparse representation of seismic data in a transform domain have been popular in the past decade because of their promising frameworks. A previous example is the project onto convex set (POCS) algorithm based on the Fourier transform method (Abma and Kabir, 2006). In recent years, several directional wavelets, including curvelets and shearlets, have been applied to sparsely present seismic events (Herrmann and Hennenfent, 2008). Yang et al. (2012) propose seismic interpolation using the curvelet transform-based POCS algorithm. These nonadaptive or highly redundant transforms have strong anisotropic directional selectivity. Considering the characteristics of seismic data, the seislet transform was presented by Fomel and Liu (2010) and later used for seismic dealiasing interpolation based on POCS (Gan et al., 2015). Dictionary learning methods (Liang et al., 2014) and rank-reduction regularization methods (Trickett et al., 2010; Gao et al., 2013a; Ma, 2013) have also been successfully applied to seismic interpolation. Yu et al. (2015) extend the data-driven tight frame (DDTF) method to 3D seismic data interpolation and later proposed the Monte Carlo DDTF method to reduce computation (Yu et al., 2016). Most of these interpolation methods are suitable only for random missing cases. For regularly subsampled seismic data with spatial aliasing, associated antialiasing techniques are included in these methods (Naghizadeh and Sacchi, 2010).

A machine learning method with support vector regression was successfully applied to seismic data interpolation by Jia and Ma (2017). Deep learning (DL), which is a fast developing branch of machine learning, has attracted significant attention from multi-disciplinary researchers. DL offers to learn an amount of parameters through the convolutional neural network (CNN) to capture high-level features in the data. Recently, DL has achieved significant progress in computer vision research, including image classification (Krizhevsky et al., 2012; He et al., 2016), denoising (Zhang et al., 2017a), and superresolution (Dong et al., 2016; Kim et al., 2016). Moreover, DL has been applied to geologic feature identification (Huang et al., 2017), seismic lithology detection (Zhang et al., 2018a), salt detection (Guillen et al., 2015; Wang et al., 2018a), and velocity inversion (Wang et al., 2018b). For seismic interpolation, primary attempts were made by Wang et al. (2019) using a residual network (He et al., 2016) and by Alwon (2018) using generative adversarial networks (Goodfellow et al., 2014) to recover seismic data from regularly subsampled observations. These end-to-end DL approaches directly learn interpolation in certain missing cases of synthetic seismic training data because of a lack of training labeled data. The testing of these approaches, however, requires feature similarity of the testing data to the training data set, which prevents the practical application of these DL methods in field seismic data processing.

Figure 1. Flowchart of the proposed CNN-POCS method. The CNN denoisers are learnt from the natural image data set rather than the seismic labeled data.

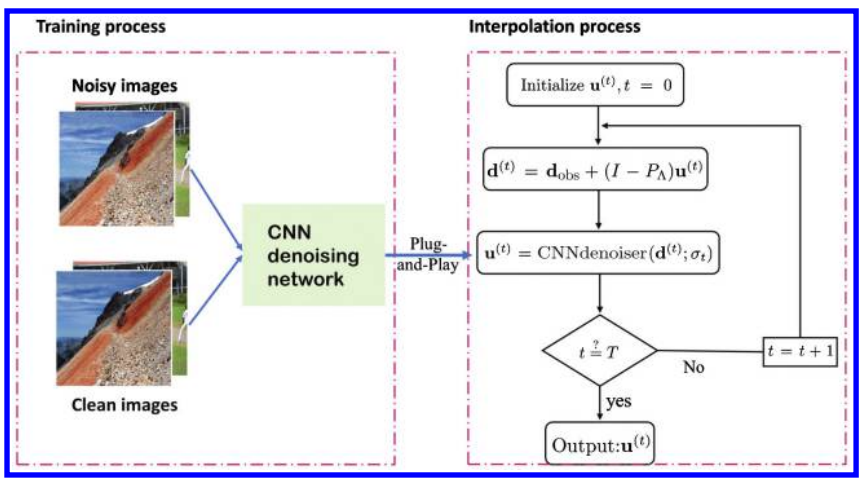

Figure 2. A simple three-layered CNN denoiser network. The input is convolved with a set of filters, as shown in red above (the sliding window), to obtain a set of feature maps.

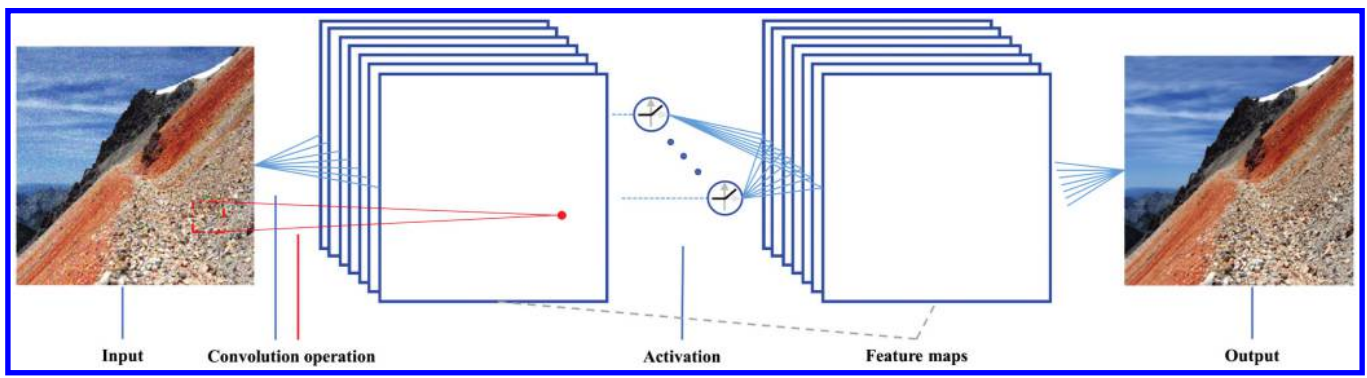

Figure 3. The architecture of the CNN denoiser used in our study: s-DConv, s-dilated convolution; BN, batch normalization (Ioffe and Szegedy, 2015); and ReLU, rectified linear units.

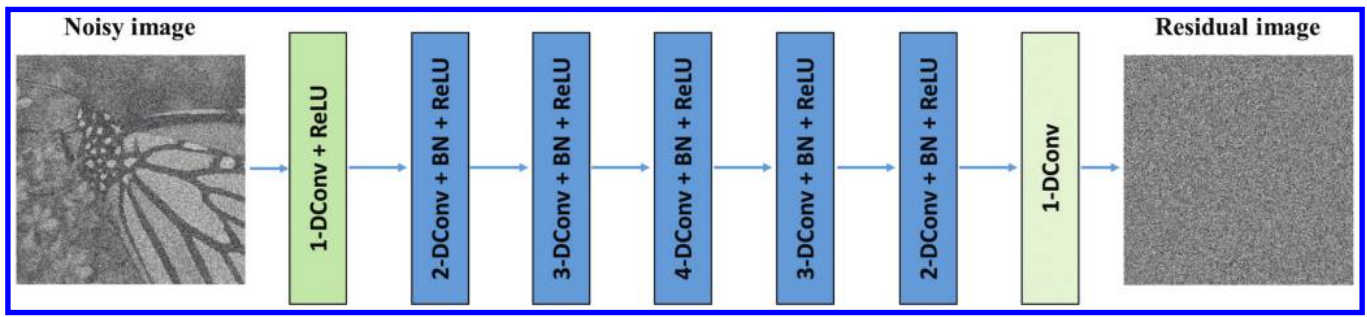

Figure 4  Dilated convolution with $3 \times 3$ nonzero entries (the red dots) in the filter. The 1-dilated convolution is equivalent to the normal convolution operation. The RF is shown.

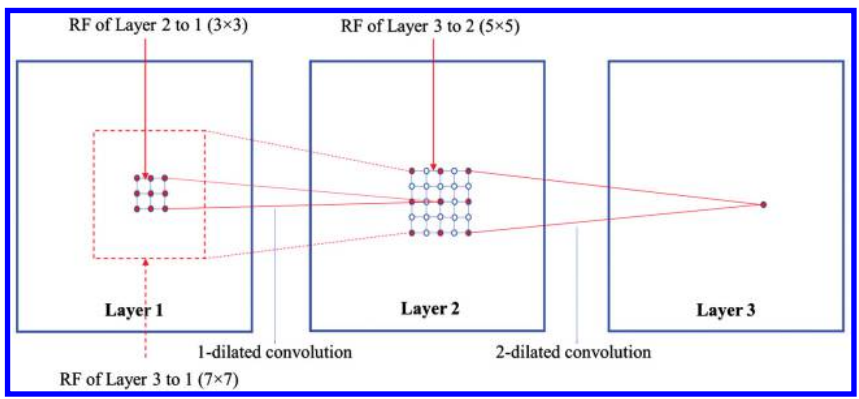

In this paper, we propose a simple and efficient approach for seismic data interpolation. The main idea is to integrate the deep denoising networks that learn denoising from natural images into the POCS algorithm. The motivation comes from the intrinsic denoising component of the POCS algorithm and the high performance of deep networks in image denoising. This study is similar in spirit to the studies using the DL network as a regularizer in image processing (Zhang et al., 2017b; Liu et al., 2018). However, whereas they used half-quadratic splitting (Geman and Yang, 1995) or alternating direction method of multipliers (ADMM) (Boyd et al., 2011) to separate the regularization term from the fidelity term and then replace the regularization term by DL networks, we use DL networks to perform the denoising mission that exists in the POCS algorithm. In the network training stage, instead of learning denoising from the seismic data, the CNNs learn denoising from noisy-clean natural image pairs. In the testing stage, these pretrained CNN denoisers are input into the POCS framework to tackle the interpolation of seismic data. This approach explores a new technique, different from transfer learning (Pan and Yang, 2010), to alleviate the lack of big data for DL in certain fields. In the testing stage of seismic interpolation, we obtain better dealiasing and synthetic data reconstruction for weak events in regular missing cases than those using the $f$-$x$ prediction-based method (Spitz, 1991). For field data interpolation, we also compare the CNN-POCS method with two other state-of-the-art methods: the curvelet transform method (Candès and Donoho, 2004; Ma and Plonka, 2010) and the block-matching and 3D filtering (BM3D) method (Dabov et al., 2008) based on the POCS strategy.

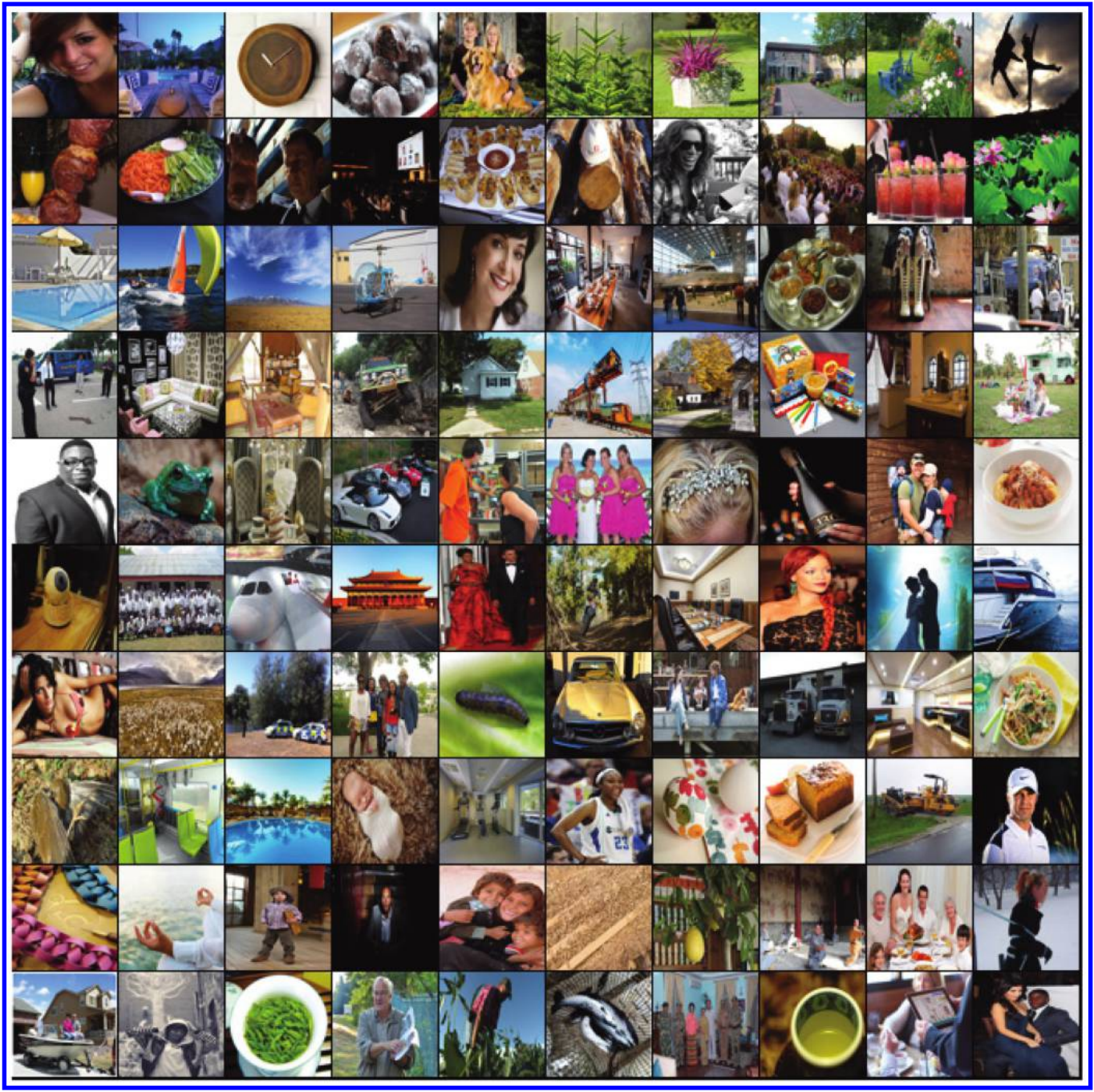

Figure 5. One hundred images randomly selected from the natural image training set.

The novelty in this study can be outlined in two aspects. (1) Unlike end-to-end DL approaches for seismic interpolation in which the networks must learn about subsampling, our method leverages the interpolation by iteratively attenuating noise using neural networks. Subsampling is not involved in learning, so it makes our method flexible and practical. (2) We observed that using neural network denoisers that learn from natural images other than seismic data could contribute toward obtaining satisfactory seismic interpolation results. This could further help to overcome the huge barrier of lacking labeled data for DL in seismic signal processing.

The rest of this paper is organized as follows. In the "Method" section, we briefly introduce the background and the POCS framework for seismic data interpolation, and then present our CNN-POCS method, the architecture of the denoising network, as well as the denoising network learning strategy. In the "Numerical experiments and results" section, we show the details of the training networks and the numerical results while testing for seismic interpolation in regular and irregular sampling cases for synthetic and field data. A discussion on the CNN-POCS method is presented in the "Discussion" section. The conclusion is presented in the final section.

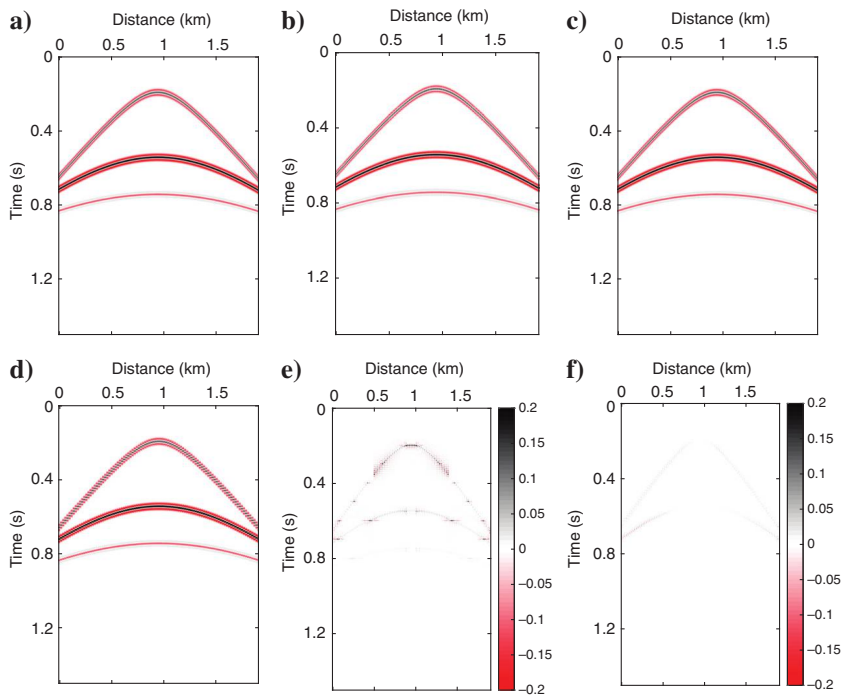

Figure 6. Interpolation of the layered model. (a) and (d) complete data and 50% regularly subsampled data with a 20 m trace interval. (b) and (e) interpolated data from the $f$-$x$ method (S/N = 26.38) and the residual. (c) and (f) interpolated data from the CNN-POCS method (S/N = 31.72) and the residual.

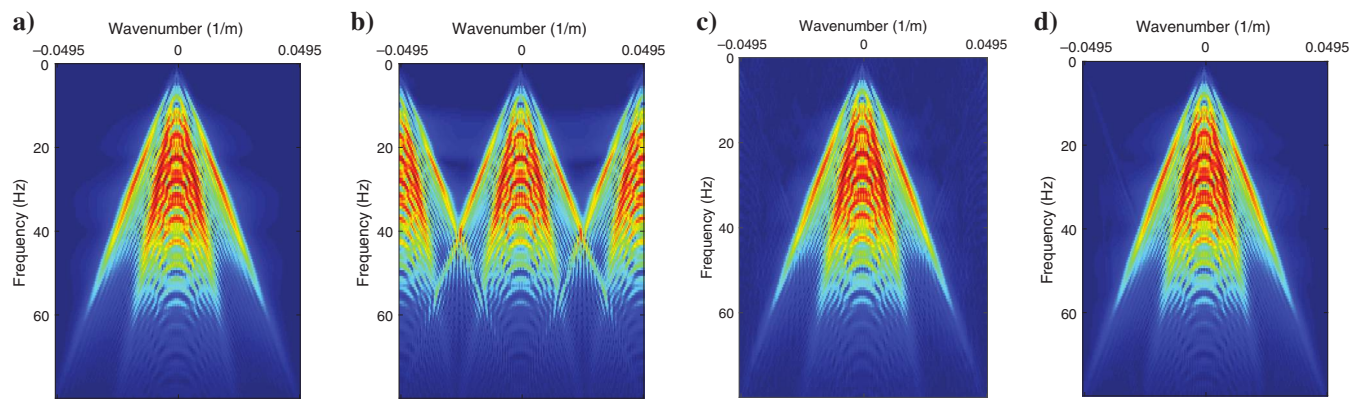

Figure 7. The $f$-$k$ spectra of the layered model. (a) Complete data, (b) regularly sampled data with a 20 m trace interval, (c) interpolated data using the $f$-$x$ prediction-based method, and (d) interpolated data using the proposed CNN-POCS method.

## METHOD

### Background and the POCS framework

Seismic data interpolation aimed at recovering the complete data $\mathbf{d}$ from an observed incomplete data $\mathbf{d}_{obs}$ can be characterized as

$$\mathbf{d}_{obs} = P_{\Lambda}\mathbf{d}, \qquad (1)$$

where $P_{\Lambda}$ denotes the subsampling matrix. Seismic data can be sparsely represented by

$$\mathbf{d} = \Phi\mathbf{x}, \qquad (2)$$

where $\Phi$ is a sparse transform, for example, a curvelet transform or a learned dictionary, and $\mathbf{x}$ is a vector of representation coefficients. Thus, we can recover the complete or dense data $\mathbf{d}$ by regularizing $\mathbf{x}$ to be sparse, that is, resolving the following optimization problem:

$$\min_{\mathbf{x}} \|\mathbf{d}_{obs} - P_{\Lambda}\Phi\mathbf{x}\|_2^2 + \lambda\|\mathbf{x}\|_1. \qquad (3)$$

This problem is often called sparsity-promoting compressed sensing reconstruction. There are many algorithms to solve this optimization problem, such as the well-known iterative shrinkage-thresholding (IST) algorithm (Daubechies et al., 2004), its acceler-ated version, the fast iterative shrinkage-thresholding (FIST) algorithm (Beck and Teboulle, 2009), and the split Bregman method (Goldstein and Osher, 2009).

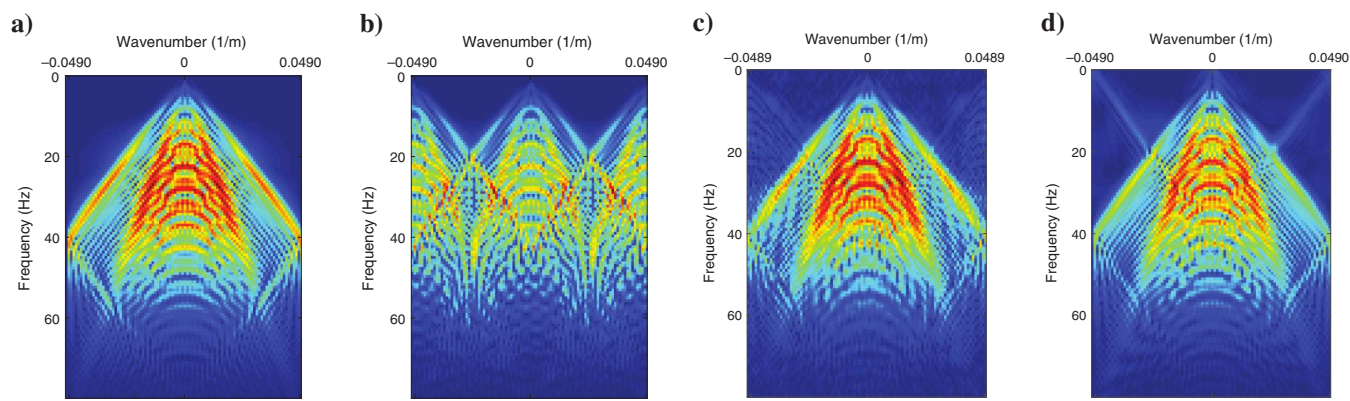

Figure 8. The $f$-$k$ spectra of the layered model. (a) Regularly sampled data with a 20 m trace interval, (b) with a 40 m trace interval, (c) interpolated data from (b) using the $f$-$x$ prediction-based method, and (d) interpolated data from (b) using the proposed CNN-POCS method. From the spectral quality point of view, the CNN-POCS method outperforms the $f$-$x$ prediction-based method except for some unexpected artifacts at low frequencies.

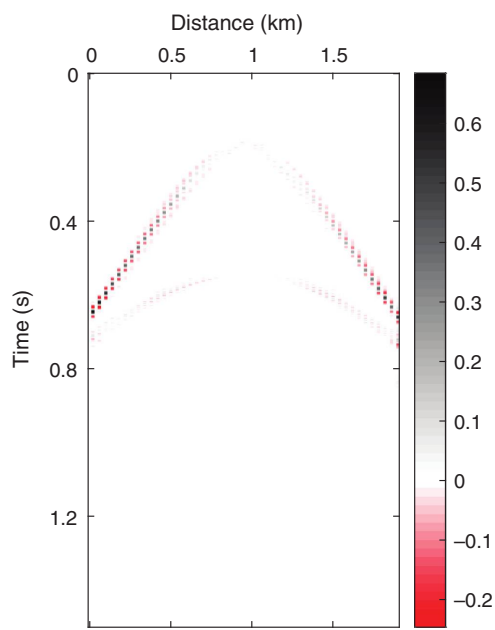

Figure 9. Reconstruction residual of the CNN-POCS method. The unexpected artifacts at low frequencies in Figure 8d correspond to the bias at the large slope region.

**Table 1. S/N (dB) comparison of the four methods on field data set 1 in regular sampling cases, a significant improvement of the CNN-POCS method over other methods when $a \leq 4$.**

| Decimating factor | $a = 5$ | $a = 4$ | $a = 3$ | $a = 2$ |
|---|---|---|---|---|
| $f$-$x$ | — | 5.93 | — | 12.15 |
| Curvelet | 3.12 | 4.29 | 6.60 | 10.00 |
| BM3D | 3.89 | 5.43 | 7.82 | 12.33 |
| CNN-POCS | 4.41 | 6.45 | 9.08 | 13.28 |

**Table 2. S/N (dB) comparison of the three methods on the field data in irregular sampling cases, a significant improvement of the CNN-POCS method over other methods when $a \geq 0.5$.**

| Sampling ratio | $a = 0.1$ | $a = 0.3$ | $a = 0.5$ | $a = 0.7$ |
|---|---|---|---|---|
| Curvelet | 18.26 | 26.54 | 37.16 | 38.03 |
| BM3D | 18.81 | 27.64 | 37.19 | 39.15 |
| CNN-POCS | 18.78 | 28.35 | 38.87 | 40.60 |

The POCS algorithm is another simple iterative method to recover **d**. It can be easily derived from the IST algorithm as follows:

$$\mathbf{u}^{(t)} = \Phi \mathcal{T}_{\lambda_t}(\Phi^T \mathbf{d}^{(t)}), \quad (4)$$

$$\mathbf{d}^{(t+1)} = \mathbf{d}_{obs} + (I - P_\Lambda)\mathbf{u}^{(t)}, \quad (5)$$

where the soft thresholding operator $\mathcal{T}_\lambda$ is defined as

$$\mathcal{T}_\lambda(x) := \begin{cases} x - \lambda \operatorname{sign}(x) & |x| \geq \lambda, \\ 0 & |x| < \lambda. \end{cases} \quad (6)$$

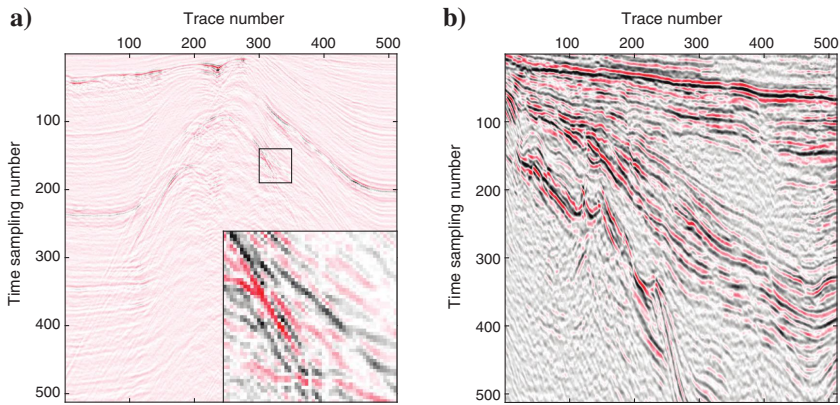

Figure 10. Two field data sets for the interpolation test. (a) Data set 1: North Sea marine data set for the regular sampling cases and (b) data set 2: data for the irregular sampling cases.

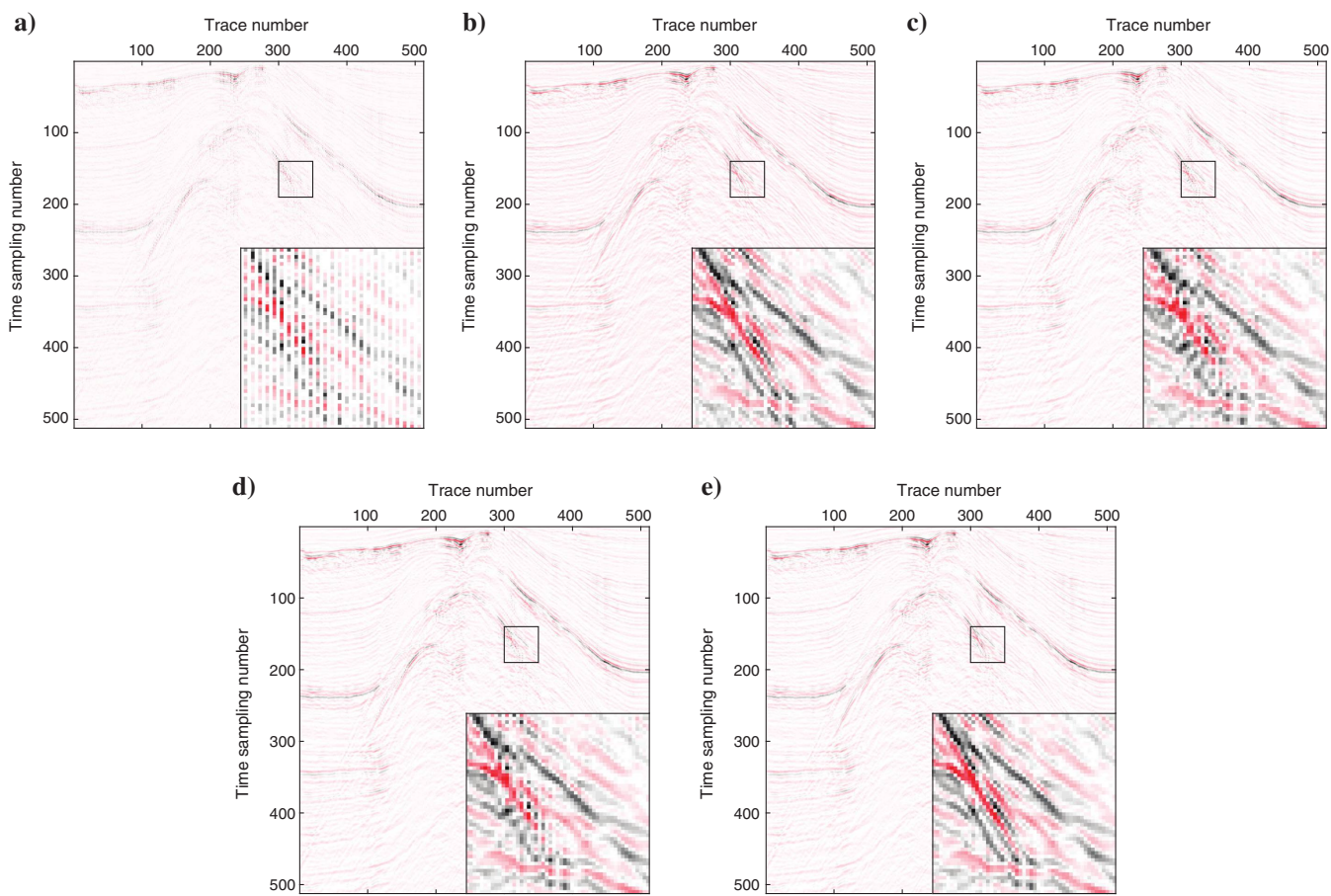

Figure 11. Restored results of the four methods on field data with a regular subsampling ratio of 0.5. (a) Subsampled data; results from the (b) $f$-$x$ method, S/N = 12.15; (c) curvelet method, S/N = 10.00; (d) BM3D method, S/N = 12.33; and (e) CNN-POCS method, S/N = 13.28.

Generally, equation 4 is regarded as a denoising procedure because the small representation coefficients, which usually correspond to noises in the signals, are eliminated during the iterations. Therefore, we can define the following POCS framework:

$$\mathbf{u}^{(t)} = \mathcal{D}_{\sigma_t}(\mathbf{d}^{(t)}), \qquad (7)$$

$$\mathbf{d}^{(t+1)} = \mathbf{d}_{\text{obs}} + (I - P_A)\mathbf{u}^{(t)}, \qquad (8)$$

where $\mathcal{D}_{\sigma_t}$ denotes the denoising operator with respect to the denoising parameter (noise variance) $\sigma_t$. We ignore the difference between the noise variance $\sigma_t$ and the thresholding parameter $\lambda_t$ of the sparse representation coefficients when $\mathcal{D}_{\sigma_t}(\mathbf{d}^{(t)}) = \Phi \mathcal{T}_{\lambda_t}(\Phi^T \mathbf{d}^{(t)})$, although $\lambda_t$ should be a function of $\sigma_t$.

The denoising operator is the key component of the POCS framework for seismic interpolation. The existing POCS algorithms for seismic interpolation, which rely on sparse representation as shown in equation 4, are actually special cases of the POCS framework. For example, the POCS algorithm based on the Fourier transform (Abma and Kabir, 2006), curvelet transform (Yang et al., 2012), dreamlet transform (Wang et al., 2014), and seislet transform (Gan et al., 2015) all perform noise attenuation by thresholding the representation coefficients in a sparse transform domain. The dictionary-based seismic interpolation methods (Yu et al., 2015; Liu et al., 2017) also fall into this scope by thresholding the dictionary sparse representation coefficients. Besides the denoisers that threshold sparse representation coefficients, the POCS framework also allows general denoisers for seismic interpolation, for example, the nonlocal means (Buades et al., 2005) and BM3D (Dabov et al., 2008) algorithms.

Another important component of the POCS framework is the scheme on the noise level $\sigma_t$. It is well-known that the POCS algorithm converges very slowly. The importance of the thresholding strategy on sparse representation coefficients was reported by Abma and Kabir (2006) when they proposed the Fourier transform POCS algorithm. Gao et al. (2010) investigate an exponentially decreasing thresholding scheme on sparse representation coefficients and obtain significant improvement on the convergence speed of the POCS algorithm. Intuitively, in our POCS framework, the noise variance $\sigma_t$ in each iteration should decrease because the recovered data progressively approximate the noise-free target. Thus, in this study, we adopt the following exponential noise level decreasing scheme for the POCS framework:

$$\sigma_t = \left(\frac{\sigma_{\min}}{\sigma_{\max}}\right)^{\frac{t-1}{T-1}} \cdot \sigma_{\max}, \quad t = 1, 2, \dots, T, \qquad (9)$$

where the two parameters $\sigma_{\min}$ and $\sigma_{\max}$ are manually tuned for seismic interpolation.

## Convolutional neural network denoiser

The POCS framework allows the use of denoisers in a general sense. A denoiser with incomparable representation capacity and denoising ability is preferred because it could potentially contribute toward improving the performance of the POCS method on seismic interpolation. Bearing this in mind and the success of DL methods in image denoising, we use the CNN as the denoiser. Unlike the linear sparse transforms with thresholding used in POCS-based algorithms mentioned above, the CNNs composed of multiple convolution operators and nonlinear activation functions such as the rectified linear unit (ReLU) (Nair and Hinton, 2010) are more non-linearly complicated and can extract features of the data in a high-level context. From the mathematical view that we provide in Appendix A, the denoising CNNs can be regarded as a set of more advanced and adaptive data-driven regularizers, compared with the sparse constraint regularizers. Thus, CNNs have an advantage over those linear sparse transforms in sparse representation and data denoising. The POCS framework associated with the CNN denoiser is summarized in Figure 1.

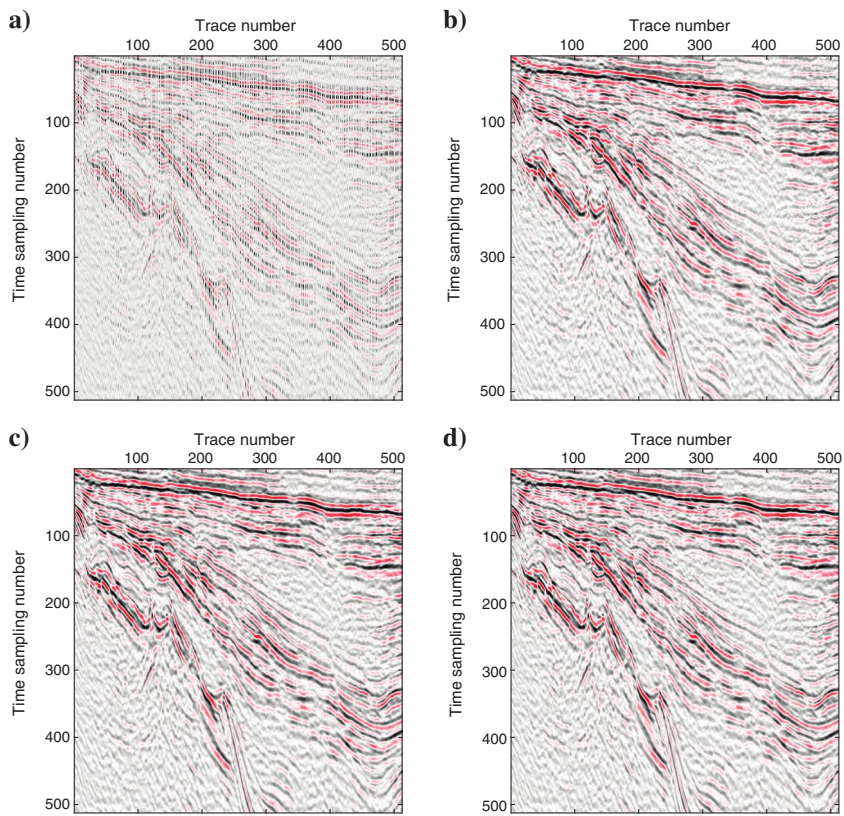

Figure 12. Restored results of the three methods on the field data with an irregular sampling ratio of 0.5. (a) Subsampled data; results from the (b) curvelet method, S/N = 37.16; (c) BM3D method, S/N = 37.19; and (d) CNN-POCS method, S/N = 38.87.

Before we detail the architecture of the denoising network used in our study, we present a basic overview of the CNN for those who are unfamiliar with it. A three-layered CNN is shown in Figure 2. The input is convolved with a set of filters to obtain a set of feature maps. The activation then introduces the nonlinearity such as the sigmoid function and ReLU. The results of the activation are further convolved with another set of filters, leading to higher level feature maps. Finally, these feature maps are convolved to obtain the output. The convolutions lead the network to detect edges and lower level features in the earlier layers and more complex features in the deep layers in the network. In supervised DL, the networks $f$ are forced to learn the filters/weights $\Theta$ by minimizing the loss function

$$\mathcal{L}(\Theta) = \frac{1}{2N} \sum_{i=1}^{N} \ell(f(\mathbf{y}_i; \Theta), \mathbf{x}_i), \tag{10}$$

through back-propagation using optimization algorithms, for example, the minibatch stochastic gradient descent (SGD) (Bottou, 2010). In equation 10, $\{(\mathbf{y}_i, \mathbf{x}_i)\}_{i=1}^{N}$ denote $N$ training pairs and $\ell$ denotes the discrepancy between the desired output and the network output.

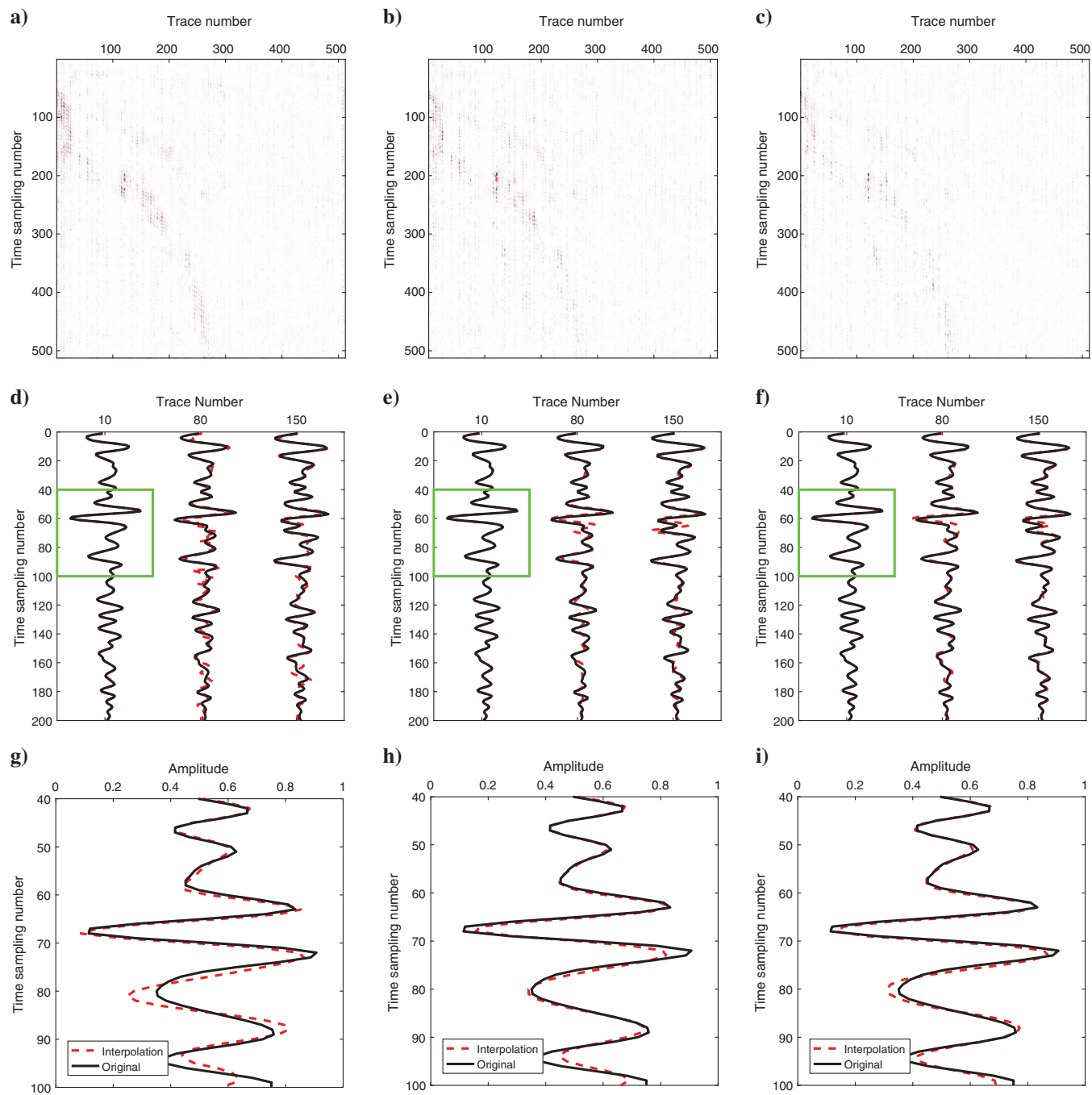

Figure 13. (a-c) Reconstruction errors. (d-f) Trace comparison. (g-i) Magnified view of the marked area in (d-f). The solid line represents the original trace, and the dotted line represents the reconstructed trace. Methods used from left to right are the curvelet method, BM3D method, and our proposed CNN-POCS method.

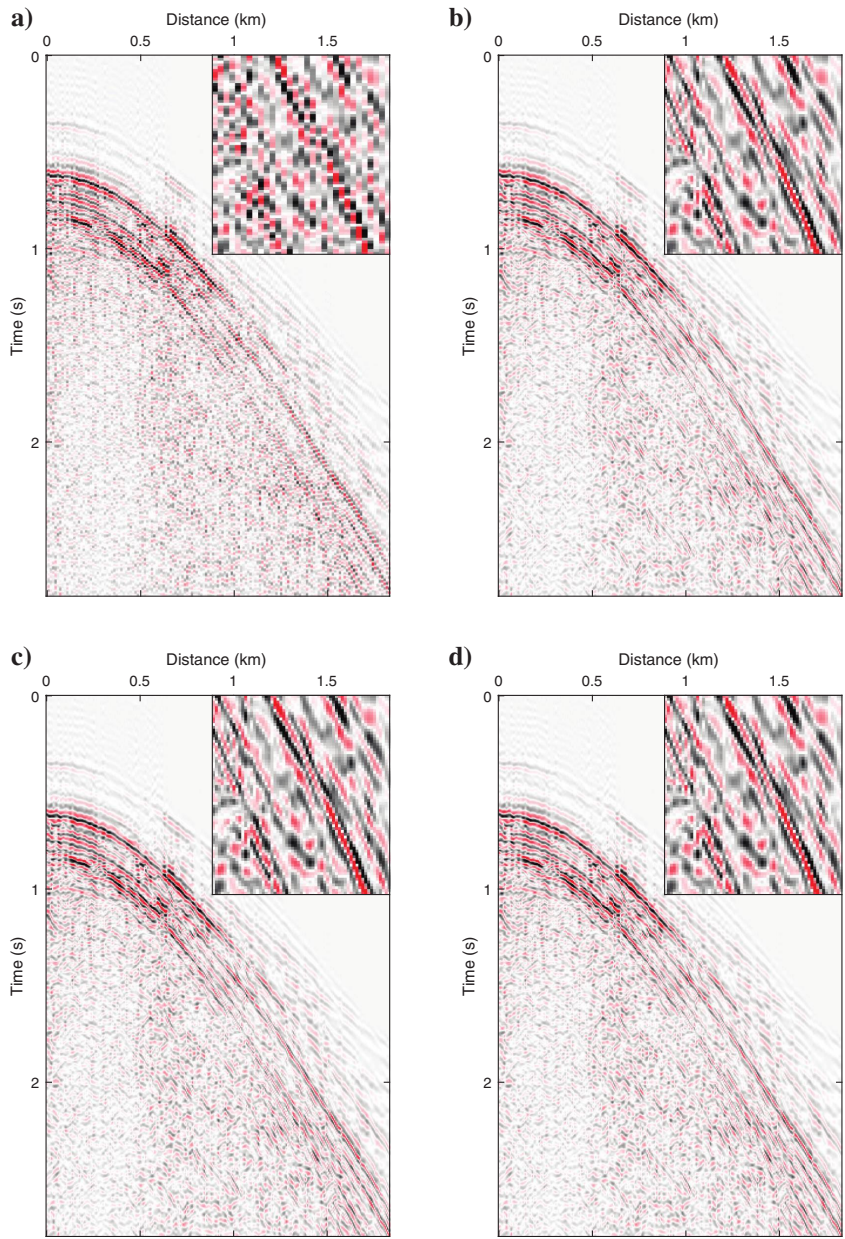

Figure 14. Dense field data reconstruction example 1. (a) Observed data with a 12.5 m trace interval, (b) reconstructed dense data with a halved trace interval, that is, 6.25 m, using our CNN-POCS method with $\sigma_{max} = 20$, (c) with $\sigma_{max} = 24$, and (d) with $\sigma_{max} = 30$. The region (1.6–1.88 s and 1.0–1.375 km) is enlarged at the top-right corner.

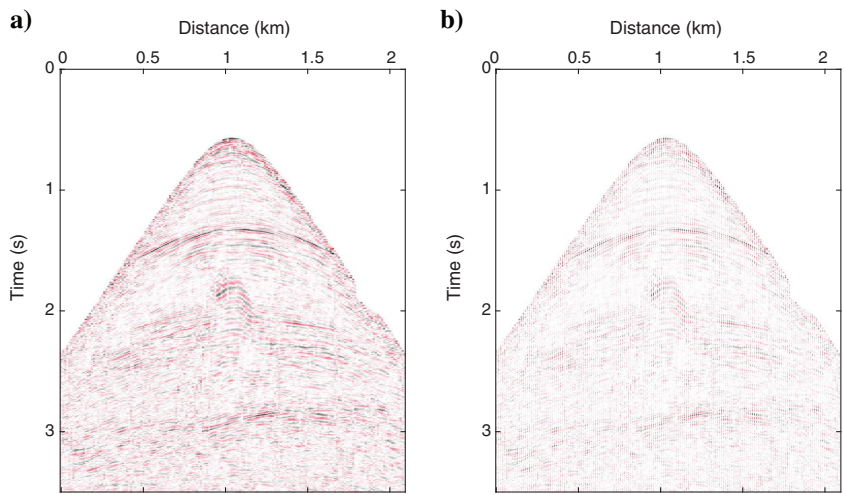

Figure 15. Dense field data reconstruction example 2. (a) Observed data with a 12.5 m trace interval and (b) zero-padded data with a 6.25 m trace interval.

### Architecture of the CNN denoiser

A few iterations are necessary to ensure that the POCS framework converges. When the pretrained CNN denoisers are input into the POCS framework to perform seismic data interpolation, the deeper the network is, the more inference computation time it consumes. Therefore, a shallow network is preferred. We adopt the architecture of the denoising CNN proposed by Zhang et al. (2017b) as illustrated in Figure 3. It consists of seven layers with three different blocks, that is, one "dilated convolution + ReLU" block in the first layer, five "dilated convolution + batch normalization + ReLU" blocks in the middle layers, and one "dilated convolution" block in the last layer. The dilation factors of the (3 × 3) dilated convolution from the first layer to the last layer are set to 1, 2, 3, 4, 3, 2, and 1. Each middle layer has 64 feature maps. The dilated convolution (Yu and Koltun, 2016) is an extension of the normal convolution, which aims to enlarge the receptive field (RF) of the networks to capture the context information while retaining the merits of normal convolution. A dilated filter with dilation factor $s$ can be interpreted as a sparse filter of size $(2s + 1) \times (2s + 1)$. Figure 4 illustrates the dilated convolution. Due to the residual learning strategy adopted in the network, we use the following loss function:

$$\mathcal{L}(\Theta) = \frac{1}{2N} \sum_{i=1}^{N} \|f(\mathbf{y}_i; \Theta) - (\mathbf{y}_i - \mathbf{x}_i)\|_F^2, \tag{11}$$

where $\{(\mathbf{y}_i, \mathbf{x}_i)\}_{i=1}^{N}$ represents $N$ noisy-clean image patch pairs.

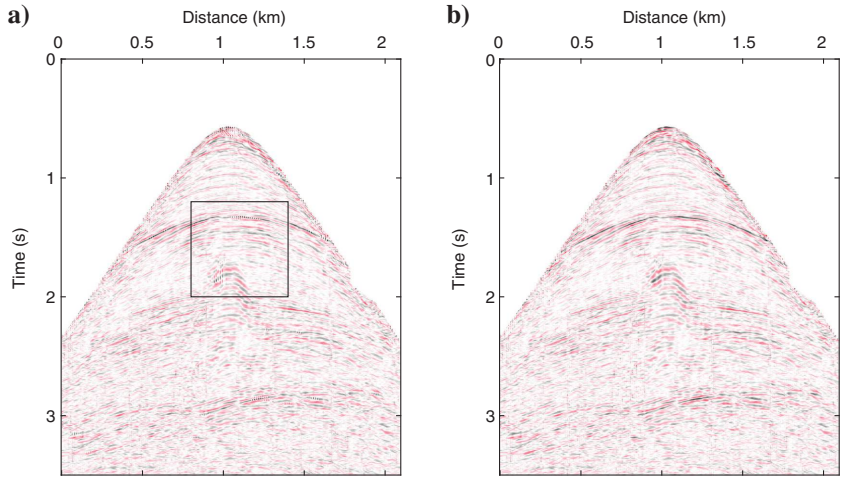

Figure 16. Dense field data reconstruction example 2. (a) Reconstruction using $\sigma_{\max} = 25$ and (b) reconstruction using $\sigma_{\max} = 50$. Some discontinuity can be observed in the rectangular region in (a), and this is suppressed in (b).

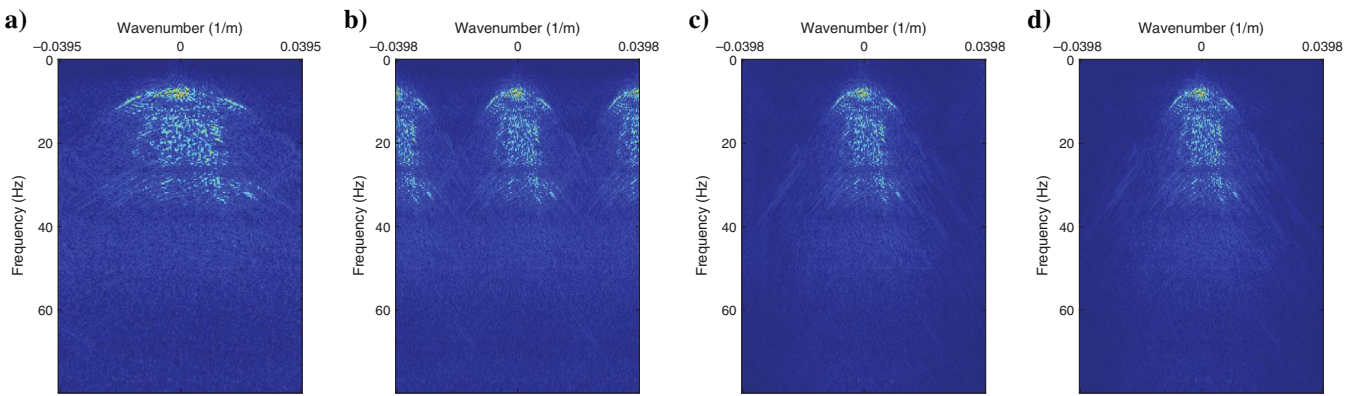

Figure 17. Spectrum comparison of the data in Figures 15 and 16. (a) Original data, (b) zero-padded original data, (c) dense reconstruction using $\sigma_{\max} = 25$, and (d) dense reconstruction using $\sigma_{\max} = 50$. Differences can be observed around the boundary between (c) and (d).

*Learning specific denoisers with small interval noise levels*

The iterative POCS framework requires various denoiser models with different noise levels; however, it is not practical to learn the CNN denoisers for all possible $\sigma_i s$. Hence, we choose to train a set of 25 denoisers in the noise-level range [0, 50] with a step size of two for each model. Another reason for this choice is that in the test stage (i.e., the interpolation on seismic data), the denoiser in the POCS framework should perform its own role regardless of the noise type and noise level of its input, which is different from recovering the latent clean image from the noisy image with an additive Gaussian noise. Thus, inexact denoising is a reasonable strategy.

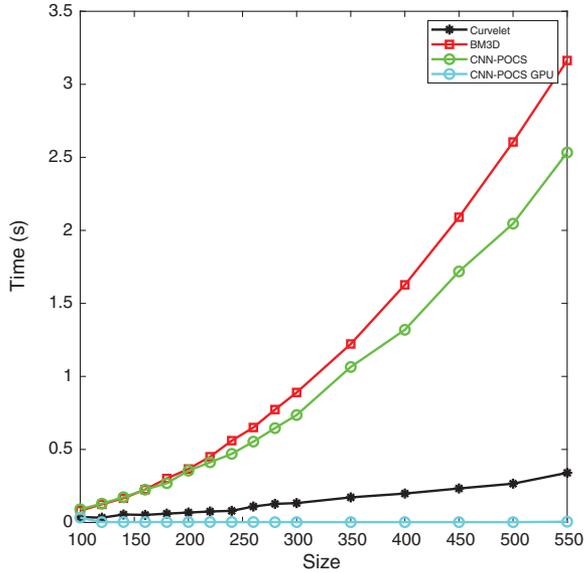

Figure 18. Running time of a single denoising step using the curvelet, BM3D, and CNN denoiser on CPU and CNN denoiser on GPU.

## NUMERICAL EXPERIMENTS AND RESULTS

In this section, we first present the data preparation and the denoising network training details. Sequentially, we use the POCS framework equipped with the pretrained denoising network to interpolate the seismic data. Interpolations for regularly and irregularly subsampled data are included in our experiments. We compare the numerical results from our method with several state-of-the-art methods, including the curvelet and BM3D methods. The traditional Spitz $f$-$x$ prediction filtering method is also used for comparison.

### Training stage: Denoisers learning from natural images

*Training data set preparation*

It is widely acknowledged that CNNs generally benefit from big training data. In seismic exploration, however, it is more difficult to obtain a large amount of input-label data pairs than in natural image processing. Although synthetic seismic data generated by wave-equation modeling can be fed into CNNs as training data, the pretrained CNNs in the testing step always require the testing data to possess feature similarity to the training data to obtain expected results (Wang et al., 2019). The feature similarity requirement essentially hinders the practical application of CNNs in seismic data processing. Given the abundance of natural images, we assume that they contain the features that are hidden in seismic data, which can be learnt by CNNs. This assumption is verified in the "Discussion" section, where the denoiser learning from images is applied to seismic data denoising. Thus, instead of using seismic data to prepare the training data set, we generate the training data set from natural images. The natural image data set used for training CNN denoiser models includes 400 Berkeley segmentation data set (BSD) images of size 180 × 180 (Chen and Pock, 2017), 400 selected images from the validation set of the ImageNet database (Krizhevsky et al., 2012), and 4744 images from the Waterloo Exploration Database (Ma et al., 2017). Figure 5 shows 100 samples drawn from this training data set. We crop the images into small patches of size 35 × 35, and the total number of patches for training is $N = 256 \times 4000$. To generate the corresponding noisy data sets, we add additive Gaussian noise to the clean patches during training.

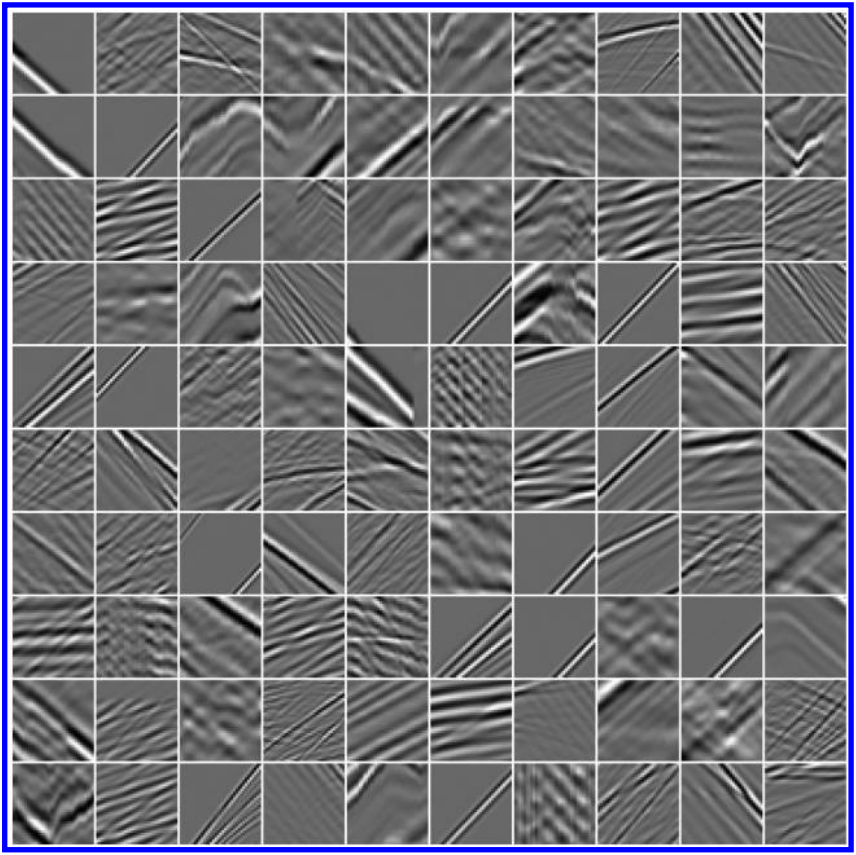

Figure 19. One hundred training examples extracted from the seismic training set.

*Training denoisers*

To optimize the network parameter $\Theta$, the Adam optimizer (Kingma and Ba, 2015) is used with the momentum parameter $\beta = 0.9$ for a minibatch size of 128. Rotation or/and flip-based data augmentation is adopted during minibatch learning. The learning rate is set to 0.001 at the start of training and then fixed at 0.0001 when the training loss stops decreasing. The training is terminated if the training loss is fixed in five sequential epochs. To reduce the overall training time, we initialize the adjacent deno-iser with the model obtained at the previous noise level. It takes about three days to finish the training of the set of denoiser models in the MATLAB (R2018a) environment with MatConvNet package (Vedaldi and Lenc, 2015) and an Nvidia Titan V GPU.

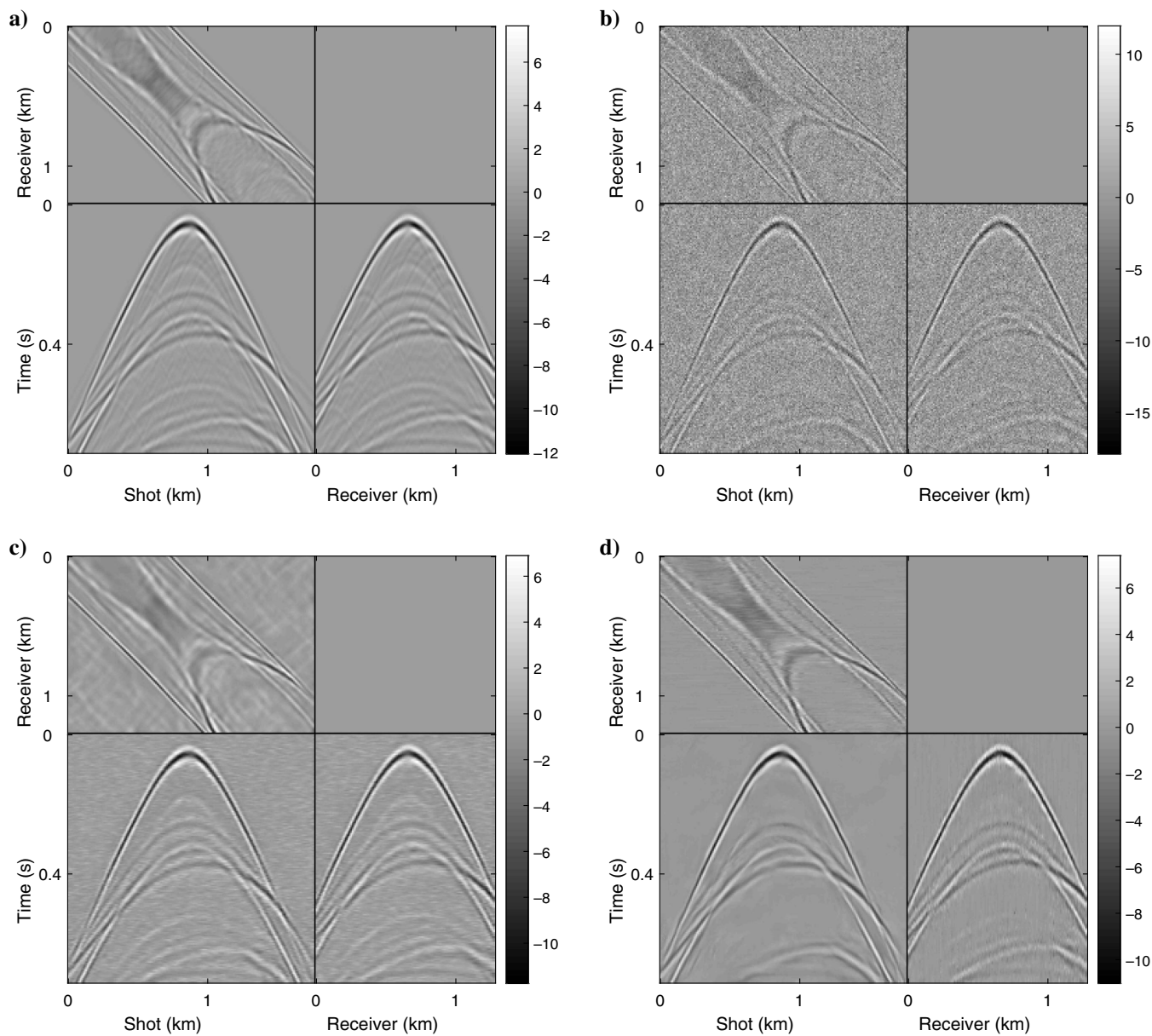

Figure 20. Three-dimensional synthetic seismic data denoising (size 178 × 178 × 128). (a) Original data, (b) noisy data (S/N = −4.21), (c) restored result by the seismic denoiser (S/N = 6.92), and (d) restored result by the image denoiser (S/N = 10.16).

### Testing stage: Seismic data interpolation

Once the denoisers are provided, we can interpolate the seismic data by the POCS algorithm. We consider the interpolation results from the curvelet transform method, BM3D method, and the $f$-$x$ prediction filtering method for comparison with our CNN-POCS method. We fix the number of iterations $T$ of the POCS algorithm at $T = 30$ for all of the denoising methods, that is, the curvelet, BM3D, and CNN denoiser methods. The parameters for interpolation $\sigma_{\max}$ and $\sigma_{\min}$ are adapted to obtain the best results possible for each experiment. In practice, we fix $\sigma_{\min} = 2$ for our CNN-POCS method because the minimum noise level that the pretrained CNN denoisers can deal with is two. This setting eases the parameter fine-tuning of the CNN-POCS method to obtain better interpolated results.

The signal-to-noise ratio (S/N) value that is used to judge the quality of the restoration is defined as follows:

$$\text{S/N} = 10 \log_{10}\left(\frac{\|\mathbf{d}_0\|_F^2}{\|\mathbf{d}_0 - \mathbf{d}^*\|_F^2}\right), \tag{12}$$

where $\mathbf{d}_0$ and $\mathbf{d}^*$ denote the complete data and their reconstruction, respectively.

#### Interpolation for synthetic data set

First, we demonstrate the effectiveness of our method in the interpolation of synthetic data. Figure 6a shows the synthetic data with three events, which includes 191 traces with a 10 m trace interval. There are 751 time samples per trace with 2 ms as the time interval. Figure 6d shows the regularly sampled data with a 20 m trace interval, which contains many jaggies. The interpolated result using the proposed CNN-POCS method is shown in Figure 6c, with the recovered S/N equal to 31.72 dB. The reconstruction error is shown in Figure 6f, in which the reconstruction residual has a very small magnitude. For comparison, we present the reconstruction result from the $f$-$x$ method (S/N = 26.38 dB) and the residual in Figure 6b and 6e, respectively. To further assess the performance of the proposed CNN-POCS method, we provide the $f$-$k$ spectra in Figure 7a–7d, which represent the spectra of the true complete data, the 50% regularly subsampled data, the interpolated result using the $f$-$x$ prediction-based method, and the proposed CNN-POCS method, respectively. Spatial aliasing is observed in Figure 7b for the regularly sampled data, and it is well-removed by the proposed method as shown in Figure 7d, which validates the dealiasing effectiveness of our proposed method.

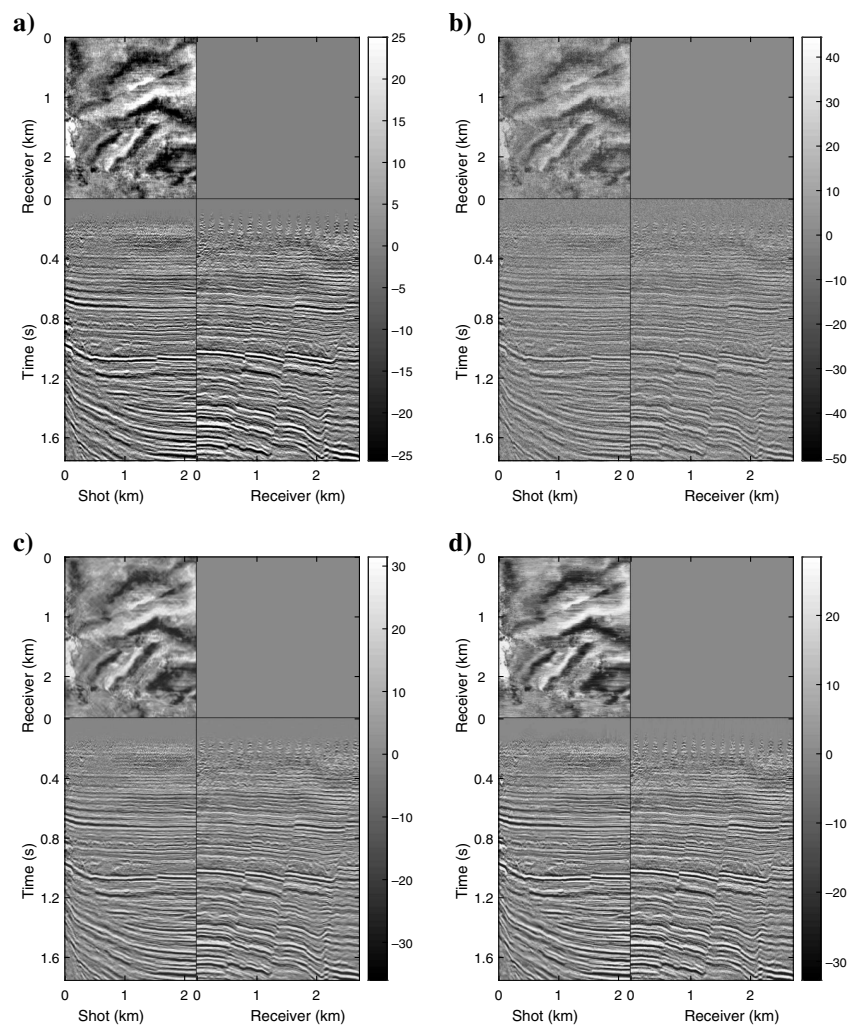

Figure 21. Three-dimensional field seismic data denoising (size $438 \times 221 \times 271$). (a) Original data, (b) noisy data (S/N = 6.99), (c) restored result by the seismic denoiser (S/N = 11.10), and (d) restored result by the image denoiser (S/N = 11.03).

We further decimate the regularly sampled data in Figure 6d with a factor of two, resulting in a 25% regularly subsampled data with more severe spectral aliasing, to assess the dealiasing effect of our proposed method. The recovered S/N are 13.98 and 13.11 dB for our proposed method and the $f$-$x$ prediction-based interpolation method, respectively. Figure 8a–8d shows the corresponding spectra. Figure 8d shows the $f$-$k$ spectrum of the interpolated data from our proposed CNN-POCS method, which outperforms that from the $f$-$x$ prediction-based method except for some unexpected artifacts at a low frequency. The reconstruction residual of the CNN-POCS method is presented in Figure 9, from where we find that the unexpected artifacts in the spectrum correspond to the reconstruction bias at the large slope and large amplitude region. These testing results further illustrate the validity of interpolating from regular sparse grids to regular and dense grids.

*Interpolation for migrated field data sets*

To further prove the flexibility of the proposed CNN-POCS method, we use the two migrated field data sets shown in Figure 10. The geologic structure has a certain amount of complexity for these two data sets. Many experiments are conducted on these two data sets, and the interpolating results are reported, with respect to regularly and irregularly subsampled data at different ratios. Tables 1 and 2 present a comparison of the S/N values for all of the sampling ratios in cases of regular sampling and irregular sampling, respectively. Here, it should be noted that irregular sampling randomly selects traces in the regular grids with a maximum trace gap equal to the inverse of the sampling ratio.

For the regular sampling cases, Figure 11 shows the results of the four methods for the subsampling ratio 50%. From the S/N value point of view, CNN-POCS produces the best result. We compare the enlarged plot of the patch marked in Figure 10a at the bottom-right portions of Figure 11b–11e. From the visual quality point of view, the interpolated result of CNN-POCS is consistent with the truth shown in Figure 10a. For the irregular sampling cases, Figure 12 shows the results of the three methods; the CNN-POCS method again obtains the best S/N value. Figure 13a–13c presents the reconstruction errors and the trace comparison of the interpolated results obtained from the curvelet, BM3D, and CNN-POCS methods are shown in Figure 13d–13i. Our proposed CNN-POCS method suffers from the fewest artifacts, and the amplitude of the artifacts is the smallest.

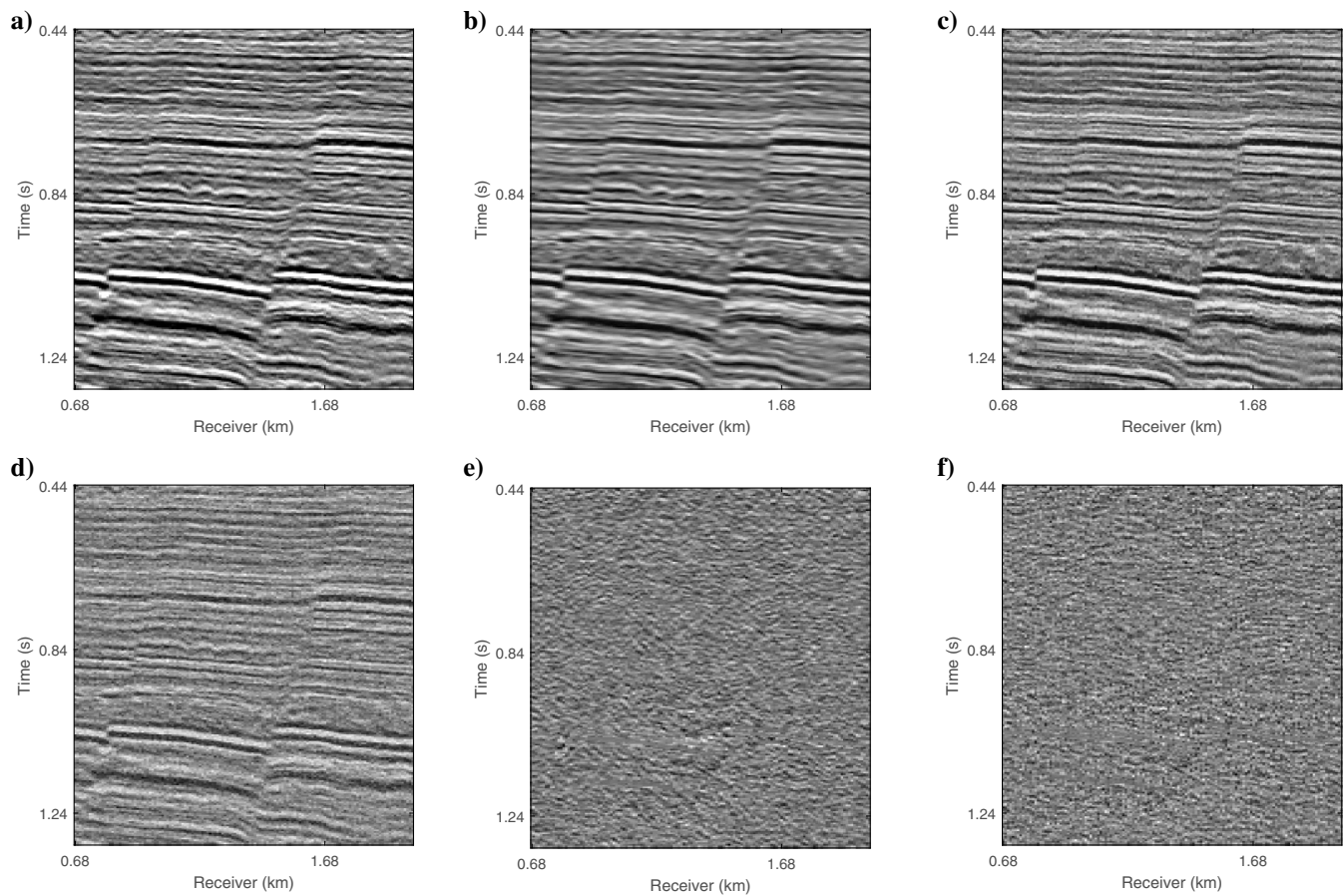

Figure 22. Magnification of a single slice of 3D field data. (a) and (d) original and the corrupted data. (b) and (e) the restored data by the seismic denoiser and the corresponding reconstruction error. (c) and (f) the restored data by the image denoiser and the corresponding reconstruction error.

### Dense field data reconstruction

In a seismic survey, sometimes the acquisition trace interval is not sufficient for a specific algorithm of indoor signal processing; thus, the interpolation algorithms should be adopted to construct dense data. Figure 14a shows a land shot gather with 147 traces, and the trace interval is 12.5 m between adjacent traces. Reconstructed dense data are provided in Figure 14b–14d with a halved trace interval using the CNN-POCS method by setting different $\sigma_{max}s$. We observe that the dense data are more continuous and the spatial serration effects are effectively weakened. A magnified version (1.6–1.88 s and 1.0–1.375 km) is presented at the top-right corner in each figure. Another dense reconstruction example is provided in Figures 15, 16, and 17. Figure 15 shows the original field data and its zero-padded version as input to the CNN-POCS algorithm. Figure 16 presents the dense reconstruction using different $\sigma_{max}$. Some gaps remain in the dense reconstruction result using $\sigma_{max} = 25$ as the rectangles in Figure 16a, and they are well-removed in Figure 16b using $\sigma_{max} = 50$. The $f$-$k$ spectrum of these data are depicted in Figure 17. The spectra aliasings in Figure 17a and 17b are well-suppressed in Figure 17c using $\sigma_{max} = 25$, but some residuals remain around the left and right boundary. These residuals are further removed by using $\sigma_{max} = 50$ as shown in Figure 17d, yielding a perfect spectral aliasing suppressed result.

### Efficiency of the proposed method

Finally, we profile the run time of one denoising step for all methods running on CPUs and a single GPU for the CNN denoiser in testing/interpolation stage in Figure 18. The experiments are conducted on a sequence of data with increasing sizes on a laptop with CPU Intel i7-9750H and a single GPU GTX 1650. The CNN denoiser takes a medium length of time on a CPU that is less than the BM3D method but more than the curvelet transform; however, it is much faster on a GPU in less than 0.005 s when the input size is 550 by 550.

## DISCUSSION

Results from the previous section show that the proposed CNN-POCS method in which the CNN denoisers are pretrained on natural images is able to produce a satisfactory interpolation quality for synthetic and field data. The data used in the tests are dominated by different features, which indicates that our proposed method does not require feature similarity among the data to be processed. Additionally, the reconstructed aliasing-free data can be beneficial to subsequent seismic data processing steps. Although the results of our study are encouraging, many questions remain to be answered further. We discuss several interesting topics arising out of our study in this section.

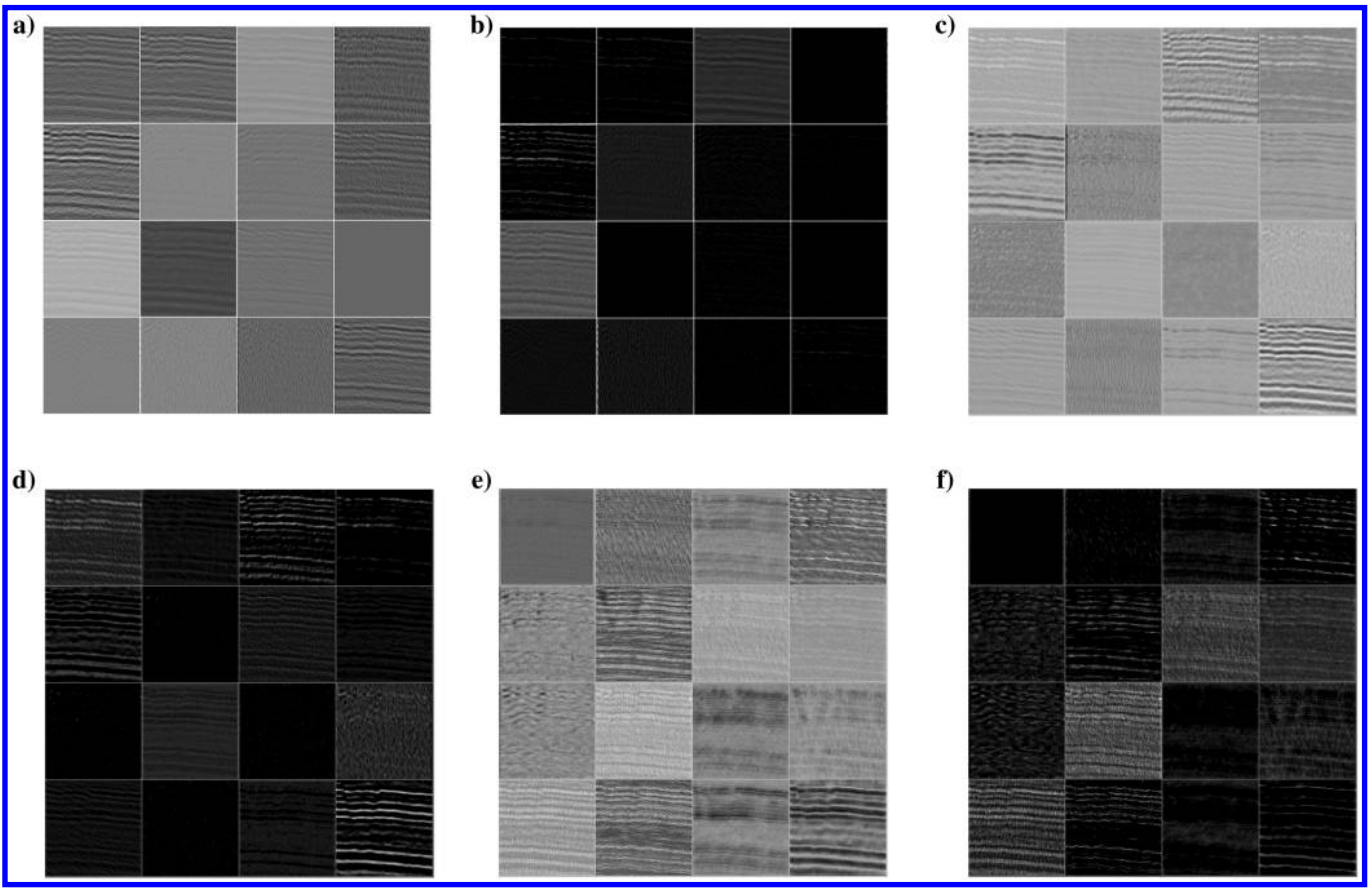

Figure 23. (a-f) Intermediate outputs of all convolutional layers in ascending order, that is, from the input to the output of the network. The outputs from the first 16 channels (out of 64) are shown in each subfigure.

## Is making the denoisers learn from images really valid for seismic data?

The cornerstone of our study is the assumption that natural images contain the features/priors of seismic data. By making the denoisers learn from natural images, we alleviate the lack of seismic labeled data. To make it convincing, we present two examples to show that the denoisers that learn from natural images (image denoisers) can get comparable performance with those that learn from seismic data (seismic denoisers) on denoising seismic noisy data. We generate 202,496 seismic samples as a training set and 84,768 seismic samples for validating from SEG open data sets. This seismic data training set is considerably large compared with that in Yu et al. (2019). One hundred patch examples drawn from this data set are shown in Figure 19. We follow the procedure of training image denoisers as described above to train the seismic denoisers. We test our image denoiser and seismic denoiser by 3D synthetic seismic data in Figure 20a and 3D field data in Figure 21a with all 2D slices denoised parallelly (inputting 3D seismic data into CNN as a batch). The corrupted data and restored results are shown in Figure 20b–20d and 21b–21d. For synthetic data, the seismic denoiser seems to fail to restore the data because of the inadequate capability of generalization, with visible noise left in the restored data, whereas our image denoiser succeeds in removing most noise and achieving S/Ns that are 3 db larger than those of the seismic denoiser. For the poststacked field data, although the image denoiser obtains an S/N value slightly smaller than that of seismic denoiser, we find that it can better preserve signals than the seismic denoiser as the reconstruction and residual magnifications depicted in Figure 22.

We further provide the intermediate outputs of all layers in the image denoising network when denoising the seismic data in Figure 23 for convolutional layers and in Figure 24 for ReLU layers from the input to the output of the network. In each subfigure, the outputs from the first 16 channels (out of 64) are presented. We observe how the seismic data feature is separated from the noise and removed step by step, and, finally, the random noise is retained (with the residual learning that we use noise as outputs).

## Is the CNN-POCS method convergent?

One may be strict with the convergency of the proposed CNN-POCS method and its sensibility to the parameters $\sigma_{max}$ and $\sigma_{min}$. We give the conditions that ensure the convergence of the POCS framework and a short proof in Appendix B, showing that the convergence requires that the denoisers are bounded and the $\sigma_t$ takes 0 as a limit. However, it is not easy to prove that the CNN denoisers we use are bounded. We do provide some experimental clues in Figure 25, which presents the footprints (the reconstructed S/N values) of the CNN-POCS method in the iterations in different interpolating cases. In each of these cases, the reconstructed S/N value lines of a large range of $\sigma_{max}$ finally merge. These results, to some extent, indicate that the CNN denoisers are bounded denoisers and the proposed CNN-POCS method is convergent. We refer readers who are interested in this part to a more recent paper by Ryu et al. (2019), which proposes the real spectral normalization to make the network strictly satisfy Lipschitz condition with Lipschitz constant smaller than one. This Lipschitz denoiser condition is indeed stronger than the bounded denoiser condition. The theoretical proof shows that the parameter $\sigma_{min}$ in our noise level exponentially decaying strategy should be positively small enough to make the CNN-POCS convergent, but in practice we find that fixing $\sigma_{min} = 2$ always obtains the best results. The reason for this is that the CNN denoisers learn from data at a minimum discrete noise level of two. For irregularly downsampled data with large data holes, slightly increasing the $\sigma_{min}$ can lead to slightly better S/N results.

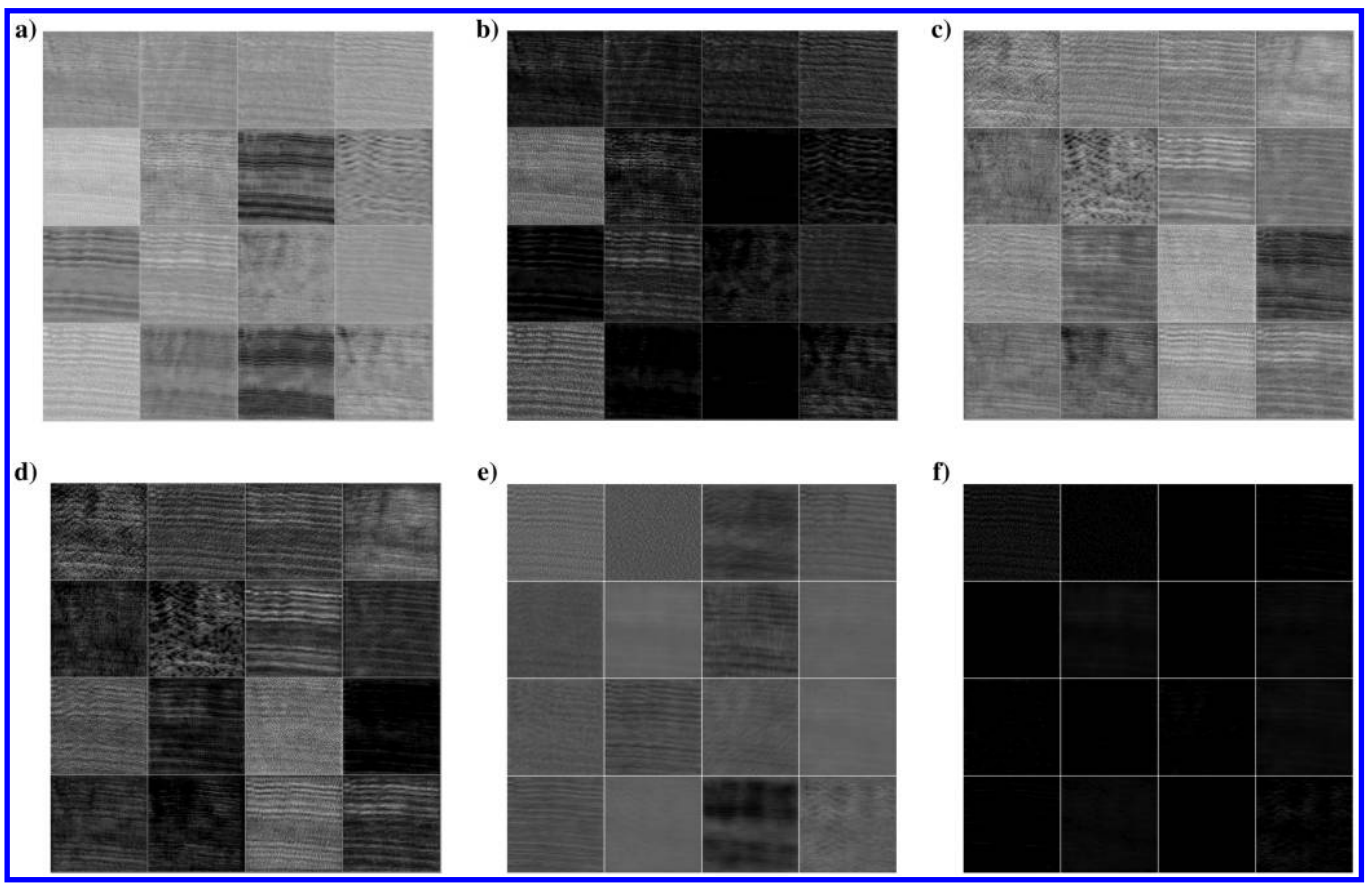

Figure 24. (a-f) Intermediate outputs of all ReLU layers in ascending order, that is, from the input to the output of the network. The outputs from the first 16 channels (out of 64) are shown in each subfigure. The events and noise are gradually separated. Black pixels indicate zeros, and gray/white pixels indicate positive values.

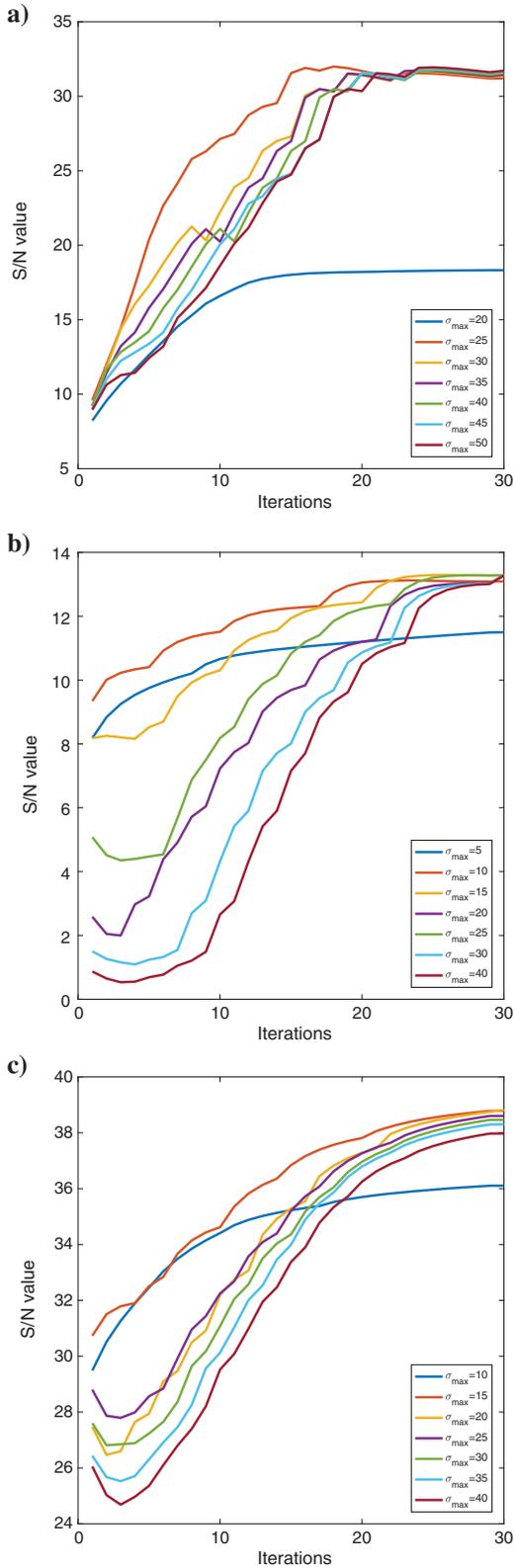

Figure 25. Reconstructed S/N values using the CNN-POCS method with different $\sigma_{max}$ in the iterations in different cases. (a) 50% regular subsampling on synthetic seismic data, (b) 50% regular subsampling on field data set 1, and (c) 50% irregular subsampling on field data set 2.

## Interpolation for randomly subsampled data

To assess the performance of the proposed method on randomly downsampled data in which some big holes occur as always being used to test traditional interpolation methods, we further provide the experiments on the synthetic data and the marine data in Figure 10b. Randomly subsampled synthetic hyperbolic data at a ratio of 50% is shown in Figure 26, and some big data holes can be observed obviously. The biggest gap is of eight traces and next to a gap of four traces, with only one data trace between each other. The interpolated results using the three methods, the curvelet, BM3D, and our CNN-POCS method, are presented in Figure 27a–27c along with the corresponding reconstruction bias in Figure 27d–27f. From the visual point of view, the proposed CNN-POCS method succeeds to reconstruct all of the data at big holes except the largest hole that has the steepest slope, whereas the BM3D method almost fails to restore the data at big holes. The curvelet method suffers from the Gibbs and boundary effects. In terms of S/N, the proposed CNN-POCS method gets the largest S/N value equal to 19.42 dB, which is more than 5 dB larger than that of the curvelet (S/N = 14.32 dB) and BM3D (S/N = 13.01 dB) methods. An example of reconstructing randomly downsampled synthetic data with two dipping events crossing each other is also provided. The synthetic complete data and the 50% randomly downsampled data are shown in Figure 28a and 28d. Some data holes are observed at/near the crossover points. The restored data and residuals using the curvelet method and our CNN-POCS method are shown in Figure 28b–28c and Figure 28e–28f. The S/N value of the result from the CNN-POCS method is slightly (approximately 0.5 dB) larger than that from the curvelet method. But again, we observe that the CNN-POCS method loses accuracy in restoring data at large holes especially where the slope is large. The underlying reasons causing this phenomenon are two fold: The first one is about the denoising neural network; the small kernel size and shallow network architecture used in our network cannot ensure that the network grasps adequate and useful information from data pixels around large data holes. Thus, the convolution layers (kernel size and convolution type) and the network depth should be optimized for better application in severely missing scenarios. The second potential reason is about the training data set; the image data set we used still lacks some characteristic features of seismic data. This encourages us to combine natural images and seismic data to form a richer training data set.

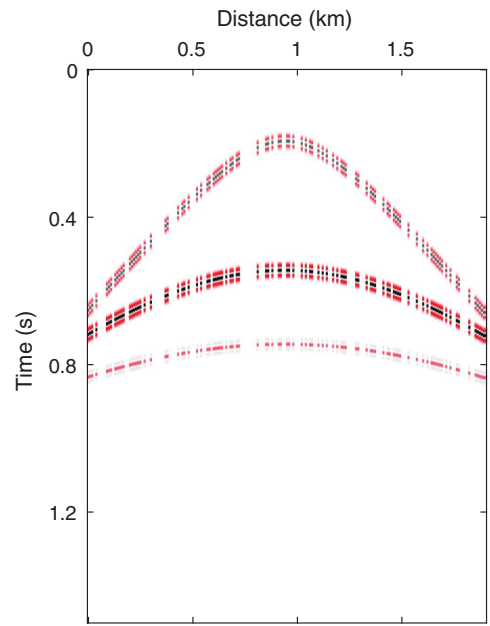

Figure 26. Randomly subsampled data of the original synthetic data in Figure 6a at a sampling ratio of 50% (S/N = 2.99). Some big data holes occur in the subsampled data.

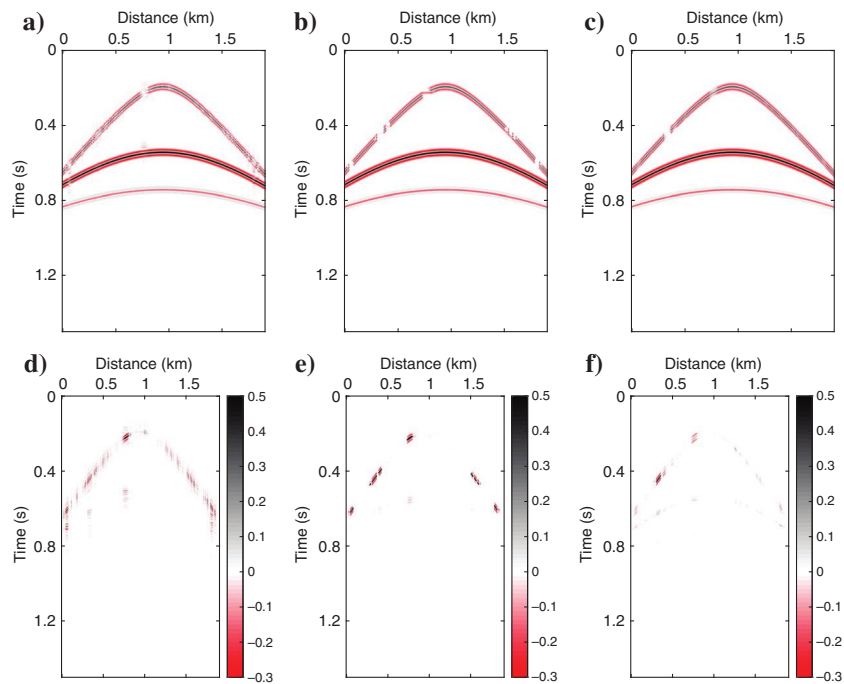

Figure 27. (a-c) Restored results by the curvelet method (S/N = 14.32), BM3D method (S/N = 13.01) and our proposed CNN-POCS method (S/N = 19.42), and (d-f) the corresponding reconstruction errors.

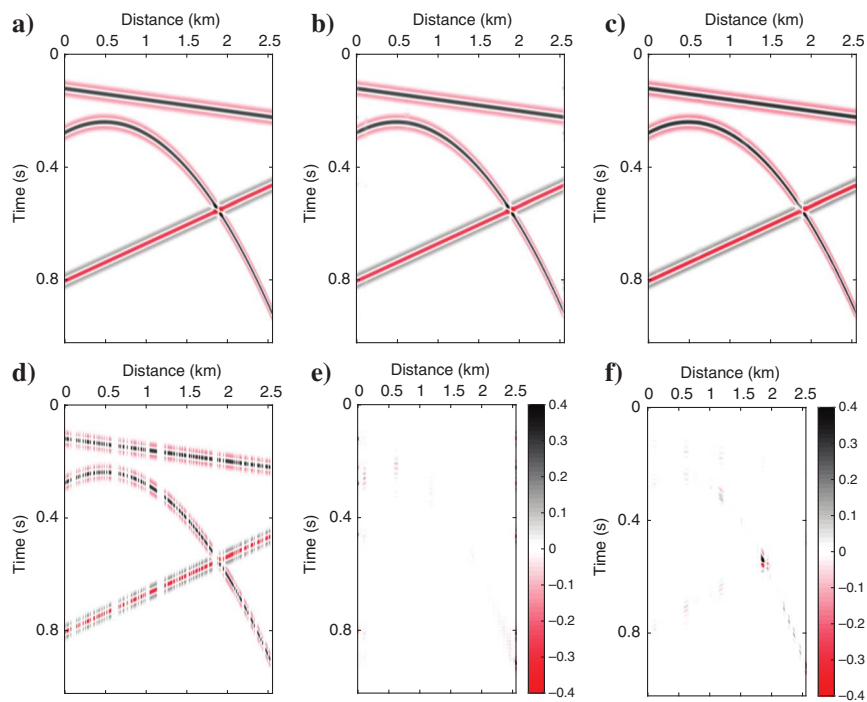

Figure 28. Experiment on synthetic data with two dipping evets crossing each other. The original range of data value is [0; 255], we scale the range of data value to [−1, 1] for presentation. (a and d) Original data and the 50% randomly down-sampled data (S/N = 19.85), large holes occur near/at cross-over points; (b and e) reconstruction by the curvelet method (S/N = 39.02) and the reconstruction residual; (c and f) reconstruction by the CNN-POCS method (S/N = 39.57) and the reconstruction residual.

Finally, we present a field data example in Figure 29, where Figure 29a shows the subsampled data and Figure 29b–29d shows the reconstruction biases of the restored results by different methods. The proposed method obtains the best S/N value equal to 32.10 dB, which is 0.6 and 1.0 dB larger than the S/N values obtained by the curvelet and BM3D methods, respectively.

## Simultaneously denoising and interpolating seismic data

The plain/original POCS framework used in our manuscript in equations 7 and 8 implicitly assumes that the observed seismic data should have a high S/N because of the insertion of the observed seismic data. However, if the observed data are noisy, this plain POCS framework will fail to suppress the noise. To solve this problem,

Gao et al. (2013b) propose the weighted strategy and then Wang et al. (2015) propose an adaptive form. However, if we review the plain POCS framework above, we will find that the noise is injected back in each iteration by the second step, although the first denoising step has attenuated the noise. Therefore, given the denoising ability of the denoiser, to make sure that the final output is noise-free, a simple yet efficient way is to switch the order of the two steps, yielding the following updated framework:

$$\mathbf{u}^{(t)} = \mathbf{d}_{\text{obs}} + (I - P_\Lambda)\mathbf{d}^{(t)}, \tag{13}$$

$$\mathbf{d}^{(t+1)} = \mathcal{D}_{\sigma_t}(\mathbf{u}^{(t)}). \tag{14}$$

If the observed data are noise free, the updated POCS framework above is equivalent to the plain POCS framework. If the observed data are noisy, say, with a noise level of $\sigma$, then to make the reconstruction well-interpolated and the noise suppressed, the only thing we have to do is to set the parameter $\sigma_{\min} = \sigma$ to make the denoiser attenuate the noise efficiently in the last iteration. An example of the synthetic hyperbolic seismic data is provided to demonstrate the efficiency of the updated POCS framework. Figure 30a presents the complete data corrupted with Gaussian noise with a noise level $\sigma = 10$ and Figure 30d presents the noisy 50% irregularly subsampled data. The reconstruction using the curvelet method and the CNN-POCS method is shown in Figure 30b–30c, and the corresponding reconstruction residual is presented in Figure 30e–30f. The missing data are well-reconstructed and the noise is well-attenuated by the CNN-POCS method, resulting in an S/N value equal to 20.24 dB, which is much larger than that the curvelet method obtains (S/N = 12.17 dB).

## Limitations

The proposed CNN-POCS method has its limitations. First, it requires multiple denoiser models to deal with different noise levels. Training such a set of denoising models is arduous. Because the noise is coupled with the truth in noisy images, the network has to learn about the noise along with image features; thus, the network cannot be adaptive to noise. It will help the network to be adaptive to noise if we provide information about noise to the network to save the effort of learning noise. The FFDNet method (Zhang et al., 2018b) feeds the network with a noise-level map and denoises the subimages to obtain a fast and flexible solution for image denoising. Thus, the burden of training multiple denoising models in the CNN-POCS method can be reduced using FFDNet. Second, the proposed CNN-POCS method loses accuracy in restoring missing seismic data at large slope and large gap regions. This is spotted in examples of interpolating severely decimated seismic data and randomly subsampled seismic data with large data holes. We discussed two main reasons for this. The first reason for limitation of neural network encourages us to optimize the architecture of denoising neural network; the second reason for the training data set encourages us to combine natural images with seismic data to establish a more powerful training data set. Third, the extension of the proposed method to 5D interpolation might be difficult if we want to directly train the networks for 5D volumes because 5D convolution is not available in those existing DL platforms. An alternative way is to train 3D CNN denoisers, which we can refer to as video denoising, and apply these 3D CNNs to the 5D seismic data along the other two axes sequentially. Or we can also apply our 2D CNN denoisers to the 5D data along the other three axes sequentially.

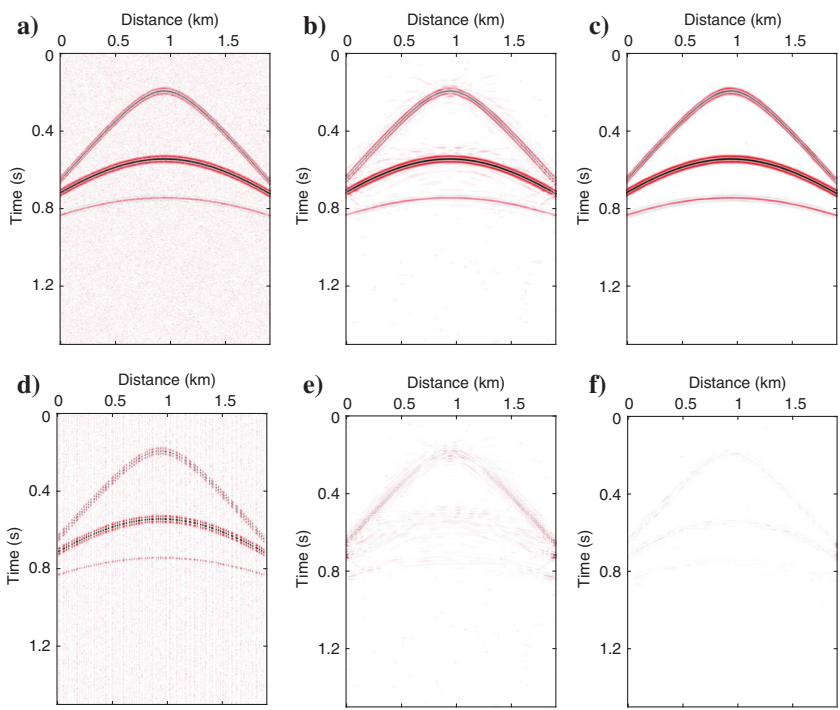

Figure 30. Simultaneously denoising and interpolating seismic data. (a) Noisy complete data (S/N = 4.80), (b) reconstruction using the curvelet method (S/N = 12.17), (c) reconstruction using the CNN-POCS method (S/N = 20.24), (d) noisy irregularly subsampled data (S/N = 1.76), (e) reconstruction residual of the curvelet method, and (f) reconstruction residual of the CNN-POCS method.

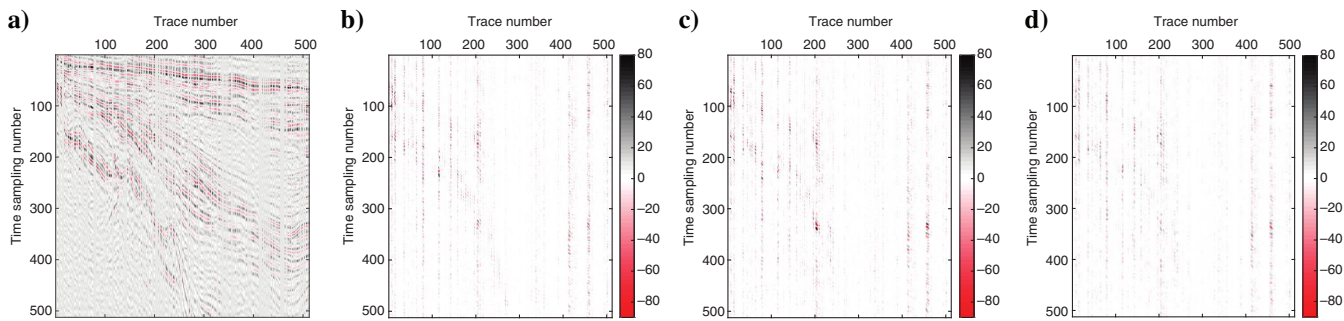

Figure 29. Randomly subsampled data and reconstruction errors using different methods. (a) Subsampled marine data with big data holes (S/N = 19.00), (b) reconstruction error of the curvelet method (restored S/N = 31.43), (c) reconstruction error of the BM3D method (restored S/N = 31.04), and (d) reconstruction error of the CNN-POCS method (restored S/N = 32.10).

## CONCLUSION

We introduced a CNN-POCS method for seismic interpolation and showed that the CNN denoisers pretrained on natural images could essentially contribute to improving the seismic interpolation results. The demand for a large amount of seismic data, as required by end-to-end deep-learning interpolation approaches, was reduced by the richness of the natural labeled images. The flexibility of our method allows it to be adaptive to any missing trace cases. Moreover, the effectiveness of the proposed method in antialiasing can be beneficial to the subsequent seismic processing steps. We tested this method on synthetic and field data, in which we considered regular and irregular sampling at different ratios. The CNN-POCS method is competitive in comparison with the $f$-$x$ method, the curvelet method, and the BM3D method in terms of S/N values and weak feature preservation. Additionally, we showed that the CNN-POCS method is stable and not sensitive to the parameters. Training the denoising models for the CNN-POCS method is slightly time-consuming: It takes approximately three days. We proposed a possible solution, that is, using a more state-of-art denoising network FFDNet, to resolve it. The denoisers that learn from natural images completely are excellent for seismic interpolation; however, they may lose some capacity to effectively represent some seismic features. We suggested a promising direction for future work, that is, mixing natural images and seismic data in the training data set for CNN denoisers. In our future work, we will further explore the extension of the idea of using plug-and-play CNNs that learn from images to seismic inversion and imaging problems.

## ACKNOWLEDGMENTS

We thank S. Yu for the very helpful discussions, and we thank Y. Sui, X. Wang, Z. Liu, and W. Wang for their help and suggestions. The work was supported in part by the National Key Research and Development Program of China under grant 2017YFB0202902 and the NSFC under grant 41625017 and grant 41804102. H. Zhang was additionally supported by the China Scholarship Council.

## DATA AND MATERIALS AVAILABILITY

Data associated with this research are available and can be obtained by contacting the corresponding author.

## APPENDIX A

## MATHEMATICAL VIEW FOR USING CNNS

Mathematically, the seismic interpolation problem can be written in a general form:

$$\min_{\mathbf{d}} g(\mathbf{d}) + \|P_\Lambda \mathbf{d} - \mathbf{d}_{\text{obs}}\|_2^2, \tag{A-1}$$

where $g$ is the prior function and a well-known example is $g(\mathbf{d}) = \|\mathbf{d}\|_1$. By the linear approximation technique, the iterative algorithm for solving this problem can be derived as

$$\mathbf{d}^{(t+1)} = \arg\min_{\mathbf{d}} g(\mathbf{d}) + <P_\Lambda \mathbf{d}^{(t)} - \mathbf{d}_{\text{obs}}, \mathbf{d} - \mathbf{d}^{(t)}> + \frac{1}{2\delta_t}\|\mathbf{d} - \mathbf{d}^{(t)}\|_2^2$$

$$= \arg\min_{\mathbf{d}} g(\mathbf{d}) + \frac{1}{2\delta_t}\|\mathbf{d} - (\mathbf{d}^{(t)} - \delta_t(P_\Lambda \mathbf{d}^{(t)} - \mathbf{d}_{\text{obs}}))\|_2^2. \tag{A-2}$$

If we denote $\mathbf{u}^{(t)} = \mathbf{d}^{(t)} - \delta_t(P_\Lambda \mathbf{d}^{(t)} - \mathbf{d}_{\text{obs}})$ then we get the algorithm

$$\mathbf{u}^{(t)} = \mathbf{d}^{(t)} - \delta_t(P_\Lambda \mathbf{d}^{(t)} - \mathbf{d}_{\text{obs}}), \tag{A-3}$$

$$\mathbf{d}^{(t+1)} = \arg\min_{\mathbf{d}} g(\mathbf{d}) + \frac{1}{2\delta_t}\|\mathbf{d} - \mathbf{u}^{(t)}\|_2^2. \tag{A-4}$$

Treating $\mathbf{u}^{(t)}$ as the "noisy" image, the second equation minimizes the residue between $\mathbf{u}^{(t)}$ and the "clean" image $\mathbf{d}$ with the prior $g(\mathbf{d})$. More precisely, according to Bayesian probability, the second equation corresponds to denoising the image $\mathbf{u}^{(t)}$ by a Gaussian denoiser with noise level $\sqrt{\delta_t}$ (Lebrun et al., 2013). If $g(\mathbf{d}) = \|\mathbf{d}\|_1$, then sparse transform can be used in this step. For unknown prior function $g(\mathbf{d})$, it encourages us to use Gaussian CNN denoisers to learn it from data.

## APPENDIX B

## CONVERGENCE CONDITION AND PROOF

Before we give the convergence condition of the POCS framework and its proof, a definition of the bounded denoiser is stated below; it will help us with our main convergence result.

DEFINITION 1. (Chan et al., 2017). Bounded denoiser: A bounded denoiser with a parameter $\sigma$ is a function $\mathcal{D}_\sigma : \mathbb{R}^n \to \mathbb{R}^n$ such that for any input $\mathbf{x} \in \mathbb{R}^n$,

$$\|\mathcal{D}_\sigma(\mathbf{x}) - \mathbf{x}\|^2 \leq n\sigma^2 C, \tag{B-1}$$

for some universal constant C independent of $n$ and $\sigma$.

The main convergence result of the POCS framework is as follows.

THEOREM 1. The POCS framework demonstrates a fixed-point convergence if the denoiser is bounded and $\sigma_t \to 0$ as $t \to \infty$. That is, there exists $\mathbf{d}^*$ such that $\|\mathbf{d}^{(t)} - \mathbf{d}^*\|_2 \to 0$.

*Proof.* To get the proof, we need to show the following fact first:

$$P_\Lambda \mathbf{d}^{(t)} = P_\Lambda \mathbf{d}_{obs} + P_\Lambda (I - P_\Lambda) \mathbf{u}^{(t-1)} = P_\Lambda \mathbf{d}_{obs} = \mathbf{d}_{obs},$$

$$\text{(B-2)}$$

with $P_\Lambda (I - P_\Lambda) = 0$ because the subsampling matrix $P_\Lambda$ is a diagonal matrix with entries being 0 or 1. Thus, we have

$$
\begin{aligned}
\|\mathbf{d}^{(t+1)} - \mathbf{d}^{(t)}\|_2^2 &= \|\mathbf{d}_{obs} + (I - P_\Lambda)\mathcal{D}_{\sigma_t}(\mathbf{d}^{(t)}) - \mathbf{d}^{(t)}\|_2^2 \\
&= \|(I - P_\Lambda)(\mathcal{D}_{\sigma_t}(\mathbf{d}^{(t)}) - \mathbf{d}^{(t)})\|_2^2 \\
&\leq \|\mathcal{D}_{\sigma_t}(\mathbf{d}^{(t)}) - \mathbf{d}^{(t)}\|_2^2 \\
&\leq n\sigma_t^2 C.
\end{aligned}
$$

$$\text{(B-3)}$$

Therefore, as $t \to \infty$, $\|\mathbf{d}^{(t+1)} - \mathbf{d}^{(t)}\|_2 \to 0$. Hence, $\{\mathbf{d}^{(t)}\}_{t=1}^\infty$ is a Cauchy sequence. Because a Cauchy sequence in $\mathbb{R}^n$ always converges, there must exist $\mathbf{d}^*$ such that $\|\mathbf{d}^{(t)} - \mathbf{d}^*\|_2 \to 0$. □